\documentclass[final,5p,times,twocolumn]{elsarticle}

\usepackage{amssymb}
\usepackage{lineno}

\usepackage{here}

\usepackage{setspace}

\usepackage{dcolumn}
\usepackage{lipsum}% http://ctan.org/pkg/lipsum
\usepackage{color}
\usepackage{ulem} % For \sout

\usepackage{overpic}

\journal{Nucl. Instrum. Meth. Phys. Res. A}

\begin{document}

\begin{frontmatter}



\title{Development of a multiwire proportional chamber \\ with good tolerance to burst hits}


\author[1,2]{N.~Teshima\corref{cor1}}
\ead{teshima@osaka-cu.ac.jp}
\author[3]{M.~Aoki}
\author[3]{Y.~Higashino}
\author[1]{H.~Ikeuchi}
\author[1]{K.~Komukai}
\author[3]{D.~Nagao}
\author[4]{Y.~Nakatsugawa}
\author[5]{H.~Natori}
\author[1,2]{Y.~Seiya}
\ead{seiya@sci.osaka-cu.ac.jp}
\author[6]{N.~M.~Truong}
\author[1,2]{K.~Yamamoto}

\cortext[cor1]{Corresponding author}

\address[1]{Osaka City University, Graduate School of Science, Osaka 558-8585, Japan}
\address[2]{Nambu Yoichiro Institute of Theoretical and Experimental Physics, Osaka 558-8585, Japan}
\address[3]{Osaka University, Graduate School of Science, Osaka 560-0043, Japan}
\address[4]{Institute of High Energy Physics (IHEP), Beijing 100-049, China}
\address[5]{High Energy Accelerator Research Organization (KEK), Ibaraki 305-0801, Japan}
%\address[6]{University of California, Davis, United States of America}
\address[6]{University of California at Davis, Department of Physics, One Shields Avenue Davis, CA 95616, USA}

\begin{abstract}
The DeeMe experiment to search for muon-to-electron conversions with a sensitivity 10--100 times better than those achieved by previous experiments is in preparation at the Japan Proton Accelerator Research Complex. The magnetic spectrometer used by the DeeMe experiment consists of an electromagnet and four multiwire proportional chambers (MWPCs). The newly developed MWPCs are operated with a high voltage (HV) switching technique and have good burst-hit tolerance. In this article, the final designs of the MWPCs, amplifiers for readout, and HV switching modules are described. Additionally, some results of MWPC performance evaluation are presented.
\end{abstract}


%% \begin{highlights}
%% \item The DeeMe experiment is planned at J-PARC MLF for muon to electron conversion search
%% \item Tracking detectors with high rate tolerance were developed
%% \item Performance measurement of the detectors was conducted with KURNS-LINAC electron beam
%% \item The position resolution meets requirements for momentum measurement at J-PARC H1 area
%% \end{highlights}


\end{frontmatter}

%\linenumbers


%==================================================
\section{Introduction} 
%==================================================
Muon-to-electron ($\mu$-$e$) conversion is one of the charged lepton flavor violation (CLFV) processes, which are strongly suppressed in the Standard Model of elementary particle physics (SM)~\cite{SM}. 
However, there are a number of theoretical models beyond the SM predicting CLFV processes with large branching ratios~\cite{BSM}. 
Therefore, an observation at a large rate should provide clear evidence of the existence of new physics.

DeeMe is an experiment to search for $\mu$-$e$ conversion in a nuclear field by using muons trapped in atomic orbits to form muonic atoms.
A signal of $\mu$-$e$ conversion is a monoenergetic 105-MeV electron emerging from a muonic atom with a delayed timing of an order of microsecond after muonic-atom formation.
The experiment is planned to be conducted at the Materials and Life Science Experimental Facility (MLF) of the Japan Proton Accelerator Research Complex (J-PARC). 
Muonic atoms are produced in a primary-proton target itself, which is hit by pulsed proton beams from the Rapid Cycling Synchrotron (RCS) of J-PARC.
To detect the electron and measure its momentum, we use a magnetic spectrometer consisting of a dipole magnet and four sets of multiwire proportional chambers (MWPCs).

In ordinary experiments searching for $\mu$-$e$ conversion, pion-production target, pion-decay and muon-transport section, and muon-stopping target are introduced to produce muonic atoms. 
However, in the DeeMe experiment, muonic atoms are directly produced in the primary pion-production target itself, which realizes a more compact and cost-effective muonic atom production. 
Nevertheless, large amounts of beam-prompt charged-particles from the primary proton-target hit the MWPCs. 
The number of charged particles hitting the detectors is estimated by simulation to be approximately $10^{8}$ particles per proton bunch with an RCS power of $1\ \mathrm{MW}$~\cite{MWPC}.
The MWPCs must detect a signal electron after exposure to such a high rate of charged particles, and it is critical to manage efficiency drop due to space-charge effects in the MWPCs.
To achieve this, gas multiplication is changed quickly to numbers of order between $1$ and $10^{4}$ by switching the high-voltage (HV) applied to the MWPCs.

The basic concept of the chamber design, the method of fast HV switching, and proof-of-principle tests using a prototype MWPC are described in~\cite{MWPC}.
In this article, the production of the final MWPCs with updated chamber design, including electrode configuration, readout amplifiers, HV switching modules, and more details of the chamber performance, are reported.

%==================================================
\section{HV-Switching Multiwire Proportional Chamber}
%==================================================
\subsection{Chamber Structure}
\begin{figure}
  \centering
  \includegraphics[width=0.4\textwidth]{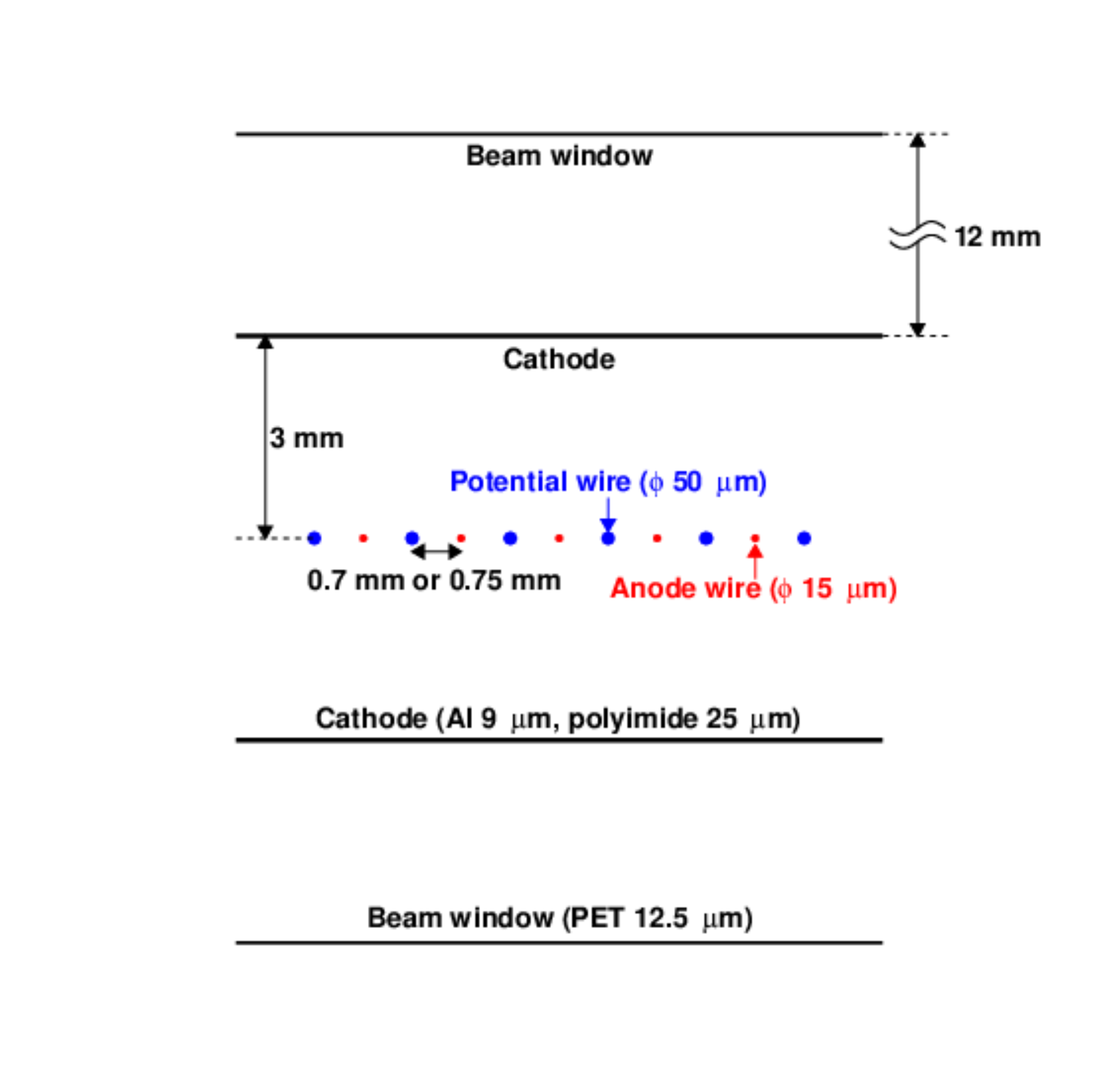}
  \caption{Schematic drawing of the wire and cathode plane configuration.}
  \label{fig:wires}
\end{figure}

Anode and potential wires are placed alternately in a center plane between two cathode planes 6 mm apart 
as shown in Fig.~\ref{fig:wires}.
Tungsten-rhenium gold-plated wires with a diameter of $15\ \mu\mathrm{m}$ are used for the anode, while tungsten gold-plated wires with a diameter of 
$50$ $\mu\rm{m}$ are used for the potential.
The wire length is approximately 300 mm.

Spacing between the anode and potential wires is 0.7 mm for two of the four MWPCs and 0.75 mm for the other two, which are denoted by 0.7-type and 0.75-type, respectively, in what follows.
The 0.7-type MWPCs of full size, having 144 anode and 145 potential wires, were manufactured first,
but they were rather unstable due to sporadic discharges.
The wire spacing of the latter two chambers was therefore widen by 0.05 mm for better discharge tolerance.
Meanwhile, the 0.7-type MWPCs were successfully operated stably
by changing the timing scheme of HV switching compared to the one adopted in~\cite{MWPC}, 
which is described later in \S\ref{sec:HV-Switching-Timing}.

The beam window is covered by a PET film with 12.5 $\mu$m thick. 
The cathode is made of an aluminum foil of 9 $\mu$m thick on a polyimide film of 25 $\mu$m thick.
The thickness of the space filled with a gas is approximately 30 mm.
The average energy loss through them is expected to be 40 keV for a minimum ionizing particle.

Because the MWPCs are operated with switching HV on potential wires, one should be careful to ensure that the resonance frequency of the wire's mechanical vibration is different from the HV switching cycle. 
A resonance frequency can be expressed as $n \sqrt{T/\rho}/2L$ ($n=1$, $2$, $\cdots$), where $L$ is the wire length, $T$ is the wire tension, and $\rho$ is the mass per unit length. %~\cite{Wire-Resonance}. 
%Anode wire tension was measured to be $(45.9 \pm 3.4)\ \mathrm{gf}$, and is now changed to $30\ \mathrm{gf}$. Assuming that the tension errors are similar, by substituting $L=300\ \mathrm{mm}$, $T=(0.29 \pm 0.03)\ \mathrm{N}$, and $\rho=3.4\times10^{-6}\ \mathrm{kg/m}$, the resonance frequencies for anode wires are $(490 \pm 30)n\ \mathrm{[Hz]}$.
For the anode wires, by substituting $L=300\ \mathrm{mm}$, $T=(0.29 \pm 0.03)\ \mathrm{N}$, and $\rho=3.4\times10^{-6}\ \mathrm{kg/m}$, the resonance frequencies estimated to be $(490 \pm 30)n\ \mathrm{[Hz]}$.
%Similarly, for the potential wires, by substituting $L=300\ \mathrm{mm}$, $T=(0.78 \pm 0.03)\ \mathrm{N}$, and $\rho=3.8\times10^{-5}\ \mathrm{kg/m}$, the resonance frequencies are obtained to be $(240 \pm 10)n\ \mathrm{[Hz]}$. 
Similarly, by substituting $L=300\ \mathrm{mm}$, $T=(0.78 \pm 0.06)\ \mathrm{N}$, and $\rho=3.8\times10^{-5}\ \mathrm{kg/m}$, the resonance frequencies for the potential wires are $(240 \pm 10)n\ \mathrm{[Hz]}$. 
The wires do not resonate when the HV switching is synchronized with the RCS cycle of $25\ \mathrm{Hz}$ because the normal frequencies of the wires are much higher.
%Therefore, the wires do not resonate when the HV switching is synchronized with the wide RCS beam cycle of $25\ \mathrm{Hz}$.

In the final design of the MWPCs, cathode planes with strip patterns are used for read out.
One of the two cathode planes is stripped into 80 channels with a width of $3\ \mathrm{mm}$ 
for measurement of the $x$ coordinate (horizontal direction), where the beam direction is defined to be the $z$ axis.
The number of read-out channels for the $y$ coordinate (vertical direction) is 16 by combining the adjacent five strips into one read-out channel.

%--------------------------------------------------
\subsection{Amplifier}\label{sect2.2}
The readout amplifiers connected to the cathode strips have 80 and 16 channels for the $x$ and $y$ axes, respectively. 
They are directly mounted on the connectors of the MWPCs. 
The outputs are sent to 100-MHz 10-bit fast ADCs~\cite{fadc} to record waveforms through long cables with a length of approximately $15\ \mathrm{m}$.

Stray capacitance between the cathode strip and potential wire exists due to the distance of $3\ \mathrm{mm}$ between them.
When the voltage on the potential wires is switched, a large current flows into the amplifier through the stray capacitance.
Therefore, the amplifier must be designed to have sufficient tolerance to large currents induced by the HV switching.

\begin{figure*}[ht]
  \centering
  \includegraphics[width=0.7\textwidth]{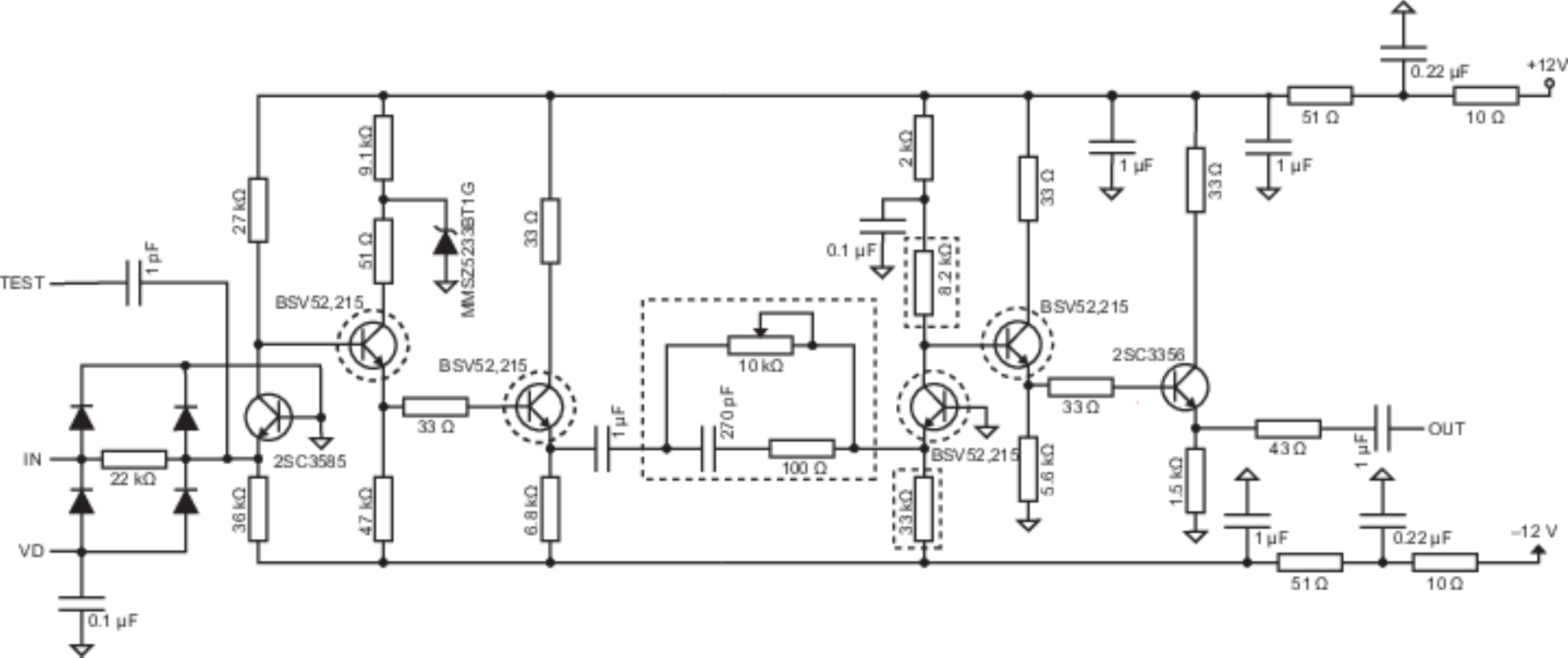}
  \includegraphics[width=0.45\textwidth]{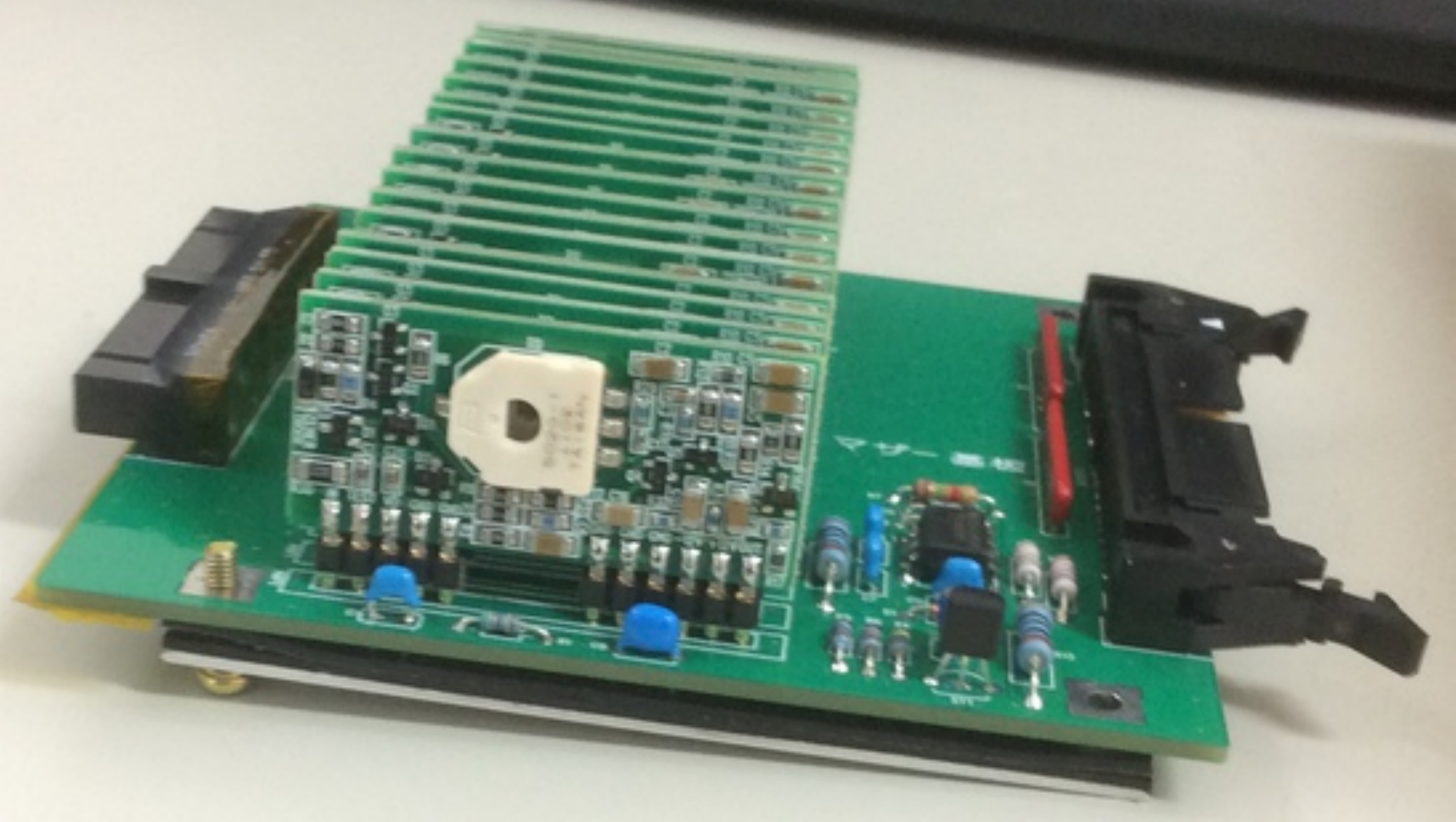}
  \caption{\label{fig:ampcir}Amplifier circuit for one channel of the MWPCs (top) and a photo of 16 channel amplifiers (bottom). The circles or boxes of dashed line in the top figure represent modified parts with respect to the original circuit of \cite{4}.}
\end{figure*}

The amplifier is modified from the readout circuit of the VENUS vertex chamber in the TRISTAN experiment at High Energy Accelerator Research Organization (KEK)~\cite{4}.
In particular, there are three points for modification: 1) to use bipolar junction transistors with more tolerance to electric currents, 
2) to increase the gain of the amplifier by changing the resistance of the second stage, 
and 3) to insert a pole-zero-cancellation circuit (PZC) to shorten the long tail of the MWPC output due to a large number of prompt charged particles~\cite{5}.
Recently, the negative range of the amplifier was increased to prevent the output waveform from saturating,
and this version of amplifier has been mass-produced (Fig.~\ref{fig:ampcir}). The modified parts are indicated in Fig.~\ref{fig:ampcir} by dashed-line circles or boxes.
%
%\begin{figure}
%  \centering
%  \includegraphics[width=0.45\textwidth]{images/ampphoto.png}
%  \caption{\label{fig:ampphoto}Photo of 16 channel amplifiers.}
%\end{figure}
%
%

%--------------------------------------------------
\subsection{HV Switching}
\begin{figure}
  \centering
  \includegraphics[width=0.45\textwidth]{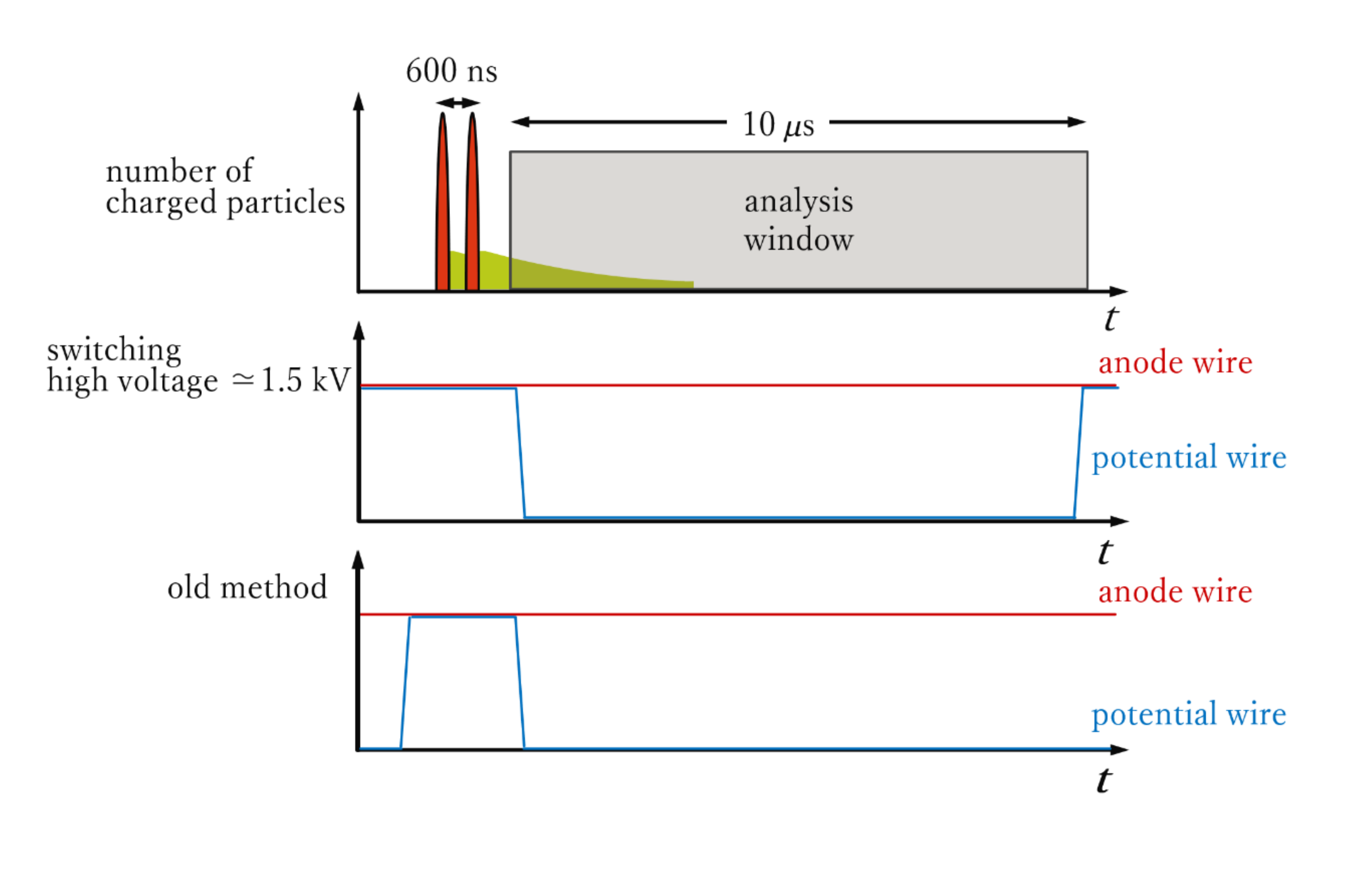}
  \caption{\label{fig:methodtoapplyHV}Schematic illustration of time structure of prompt charged particles to hit the MWPCs (top) and how the HV switching is performed (middle and bottom). The tail following the prompt pulses represents contribution of electrons produced with delayed timings. Next double pulses come after $40\ \mathrm{ms}$.}
\end{figure}

The upper part of Fig.~\ref{fig:methodtoapplyHV} schematically illustrates the time line of charged particles that will hit the detectors. 
The RCS beam has a double bunch structure, and the interval between the two bunches is $600\ \mathrm{ns}$. The repetition is $25\ \mathrm{Hz}$ so that the next double pulse comes after $40\ \mathrm{ms}$. The protons hit the target and generate prompt charged particles. The charged particles with momenta of approximately $105\ \mathrm{MeV/}c$ pass through the secondary beam line (high momentum muon beamline, H-Line~\cite{HLine}) and hit the detectors. After prompt particles hitting through, the analysis window is opened to search for a signal electron of $\mu$-$e$ conversion.
%After $800\ \mathrm{ns}$ of hitting the target, the analysis window is opened to search for an electron of $\mu$-$e$ conversion.

\subsubsection{Timing of the HV switching}\label{sec:HV-Switching-Timing}
The middle part in Fig.~\ref{fig:methodtoapplyHV} shows the scheme to apply HV for the MWPCs. 
To control the gas multiplication dynamically, the voltage on the potential wires is switched between the same HV as the one for the anode wires and 0 ${\rm V}$. The spacing between the potential and anode wires is small compared to the gap between the wire and the cathode planes; therefore, the electric field around the anode wire is determined almost entirely by the voltages applied to the potential and anode 
wires~\cite{MWPC}. When the voltages applied to the potential and anode wires are the same, the gradient of the electric potential between the two wires positioned closely is small enough to turn off the gas multiplication. Although the voltage of the cathode strips connected to the readout electronics is kept small compared to the HV to the wires, it does not result in gas multiplication due to the large distance between the wires and the cathode. Rather, it helps to sweep out electrons that are generated by prompt incident particles to prevent the avalanche charge produced by them after turning on gas multiplication.
%Rather, it helps to sweep out ions that are generated by prompt incident particles to minimize the space charge effect.
The large voltage difference induced by switching the voltage on the potential wires to 0 $\rm{V}$ creates a strong electric field around the anode wires, enabling gas multiplication.

When the voltage difference between potential and anode wires is large, attractive electrostatic forces between them also become large. 
Assuming that wires are long enough, the capacitance between the two wires per unit length $C$ is given by $C \simeq \pi \epsilon / \mathrm{ln}(s/a)$, where $\epsilon$ is the permittivity of the filling gas, 
$s$ is the wire spacing, and $a$ is the radius of the wire. 
Ignoring the difference of diameters between the anode and potential wires and substituting $\epsilon=8.85 \times 10^{-12}\ \mathrm{F/m}$, $s=0.7\ \mathrm{mm}$, and $a=7.5\ \mu\mathrm{m}$ (the anode wire radius), $C$ will be $6\ \mathrm{pF/m}$~\cite{Permittivity}. 
%%The length of wires is $300\ \mathrm{mm}$; thus, the capacitance between the two wires is approximately $2\ \mathrm{pF}$. 
%%For the case of applying $1500\ \mathrm{V}$ to anode wires and $0\ \mathrm{V}$ to potential wires, an electric charge of $2\ \mathrm{pF} \times 1500\ \mathrm{V}=3\ \mathrm{nC}$ is accumulated. 
For the case of applying $1500\ \mathrm{V}$ to anode wires and $0\ \mathrm{V}$ to potential wires, an electric charge of $6\ \mathrm{pF/m} \times 1500\ \mathrm{V}=9\ \mathrm{nC/m}$ is accumulated. 
Because the attractive force per length between two long wires is given by
$\frac{\lambda^{2}}{2 \pi \epsilon d}$, where $\lambda$ is the charge per length, $d$ is the distance between two wires, the sum of forces acting on the anode wire with a sag of $0.1\ \mathrm{mm}$ by the two adjacent potential wires at $0\ \mathrm{V}$ is approximately $0.2\ \mathrm{mN}$.
On the other hand, due to the tension of the anode wires, there is a restoring force of approximately $30\ \mathrm{g}\times 9.8\ \mathrm{m/s^{2}} \times 0.1\ \mathrm{mm}/(300\ \mathrm{mm}/2)=0.2\ \mathrm{mN}$ opposite to the direction of the wire sag. Two competing forces are on the same order and the stability of the MWPCs may be broken when the wire sag becomes larger, 
%than 0.1 mm
as the attractive electrostatic forces overcome the restoring forces.
%This is about the same as the attractive electrostatic forces between the wires, and the stability of the MWPCs may be broken.
%The dashed line in Fig.~\ref{fig:forcevssag} shows the sum of forces acting on the anode wire by the two adjacent potential wires at $0\ \mathrm{V}$ as a function of wire sag calculated by the method of images,
%where a simple geometrical configuration is considered, as illustrated in Fig.~\ref{fig:wiresag}, and sag is defined as the displacement ($\delta$) of the middle point of the anode wire from the nominal position towards one of the potential wires.
%The dotted line represents the restoring force by wire tension, which is 30 g for the anode wire.
%The magnitude of the force is estimated to be $30\ \mathrm{g}\times 9.8\ \mathrm{m/s^{2}} \times \delta/(300\ \mathrm{mm}/2)=2.0\ \mathrm{N/m}\cdot \delta$, and the direction is opposite to that of the wire sag.
%The solid line shows the sum of the electrostatic force and restoring force. 
%This graph indicates that the stability of the MWPCs may be broken when the wire sag becomes larger than $0.1\ \mathrm{mm}$, as the attractive electrostatic forces overcome the restoring forces.
When there is no voltage difference between the two wires, the position of wires should become stable due to balanced repulsive forces.
As shown in the middle plot of Fig.~\ref{fig:methodtoapplyHV}, the duration for which there is a large voltage difference between the anode and potential wires is minimized and limited to the search analysis window, on the order of 10 $\mu$s in $40\ \mathrm{ms}$, to ensure stable MWPC operation, which should be compared to the old switching scheme~\cite{MWPC} shown at the bottom of the figure. 
The current scheme of HV switching is also expected to have an advantage for stopping consecutive discharges in $\sim 10$ $\mu$s even if they occur.
\subsubsection{HV Pulser}
\begin{figure*}[ht]
  \centering
  \includegraphics[width=0.6\textwidth]{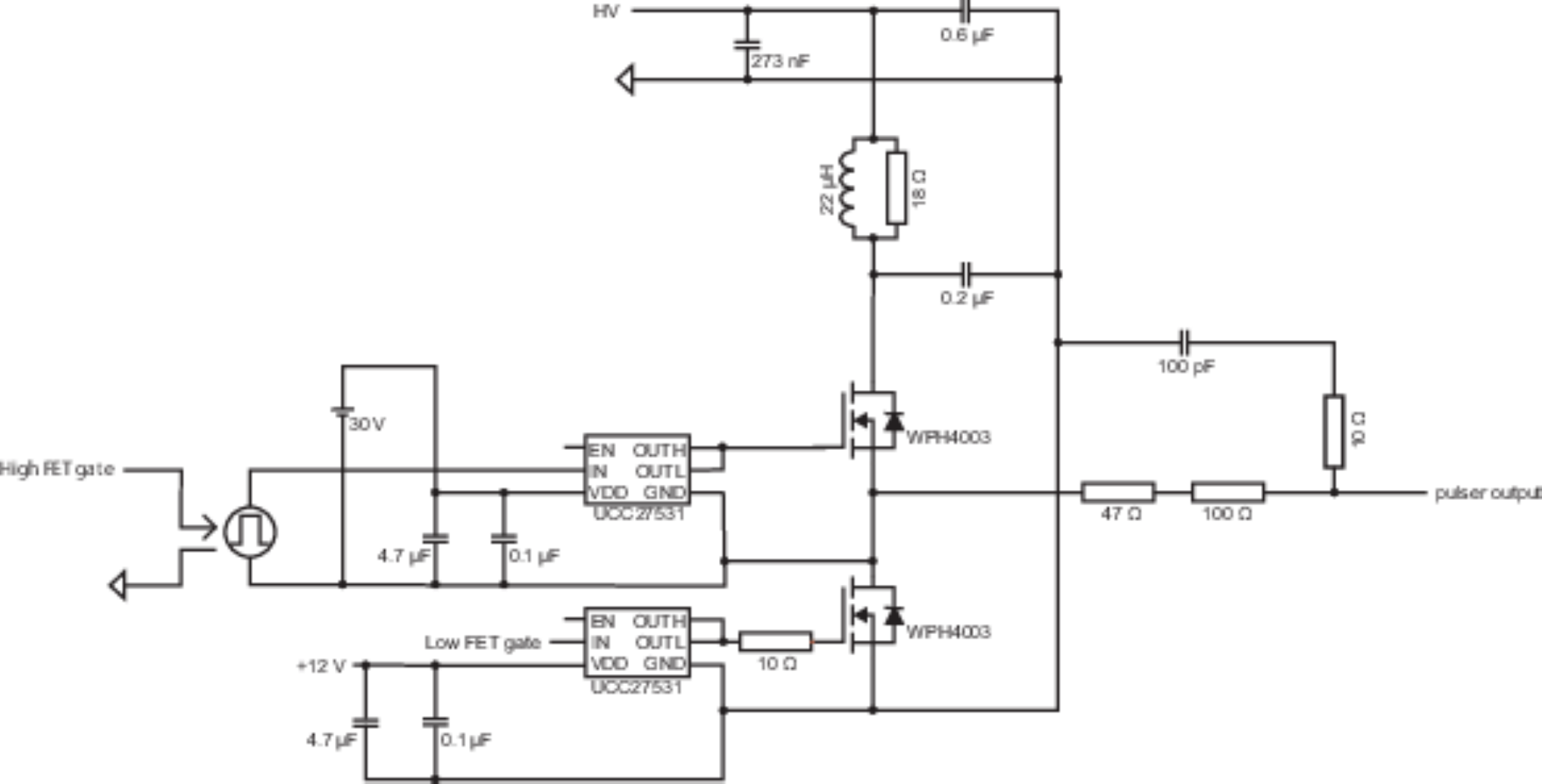}
  \caption{\label{fig:hvswitchingcircuit}Simplified circuit for the HV switching. It is inserted between the HV power supply and potential wires.}
\end{figure*}

A HV power supply 
%of the company iseg Spezialelektronik GmbH
provides DC voltages to the anode wires, while a HV switching module is inserted between the HV power supply and potential wires. 
A circuit diagram of the HV switching module is presented in Fig.~\ref{fig:hvswitchingcircuit}.

\indent
The drain of the upper MOSFET is connected to the external HV line, while the source of the lower MOSFET is grounded. 
The drain-source connection in the MOSFET is altered by the gate-source voltage. 
The circuit output is connected to the source of the upper MOSFET and the drain of the lower MOSFET. 
By controlling the drain-source connection of the two MOSFETs appropriately, 
the output voltage is switched between HV and $0 \mbox{ V}$. \\
\indent
It is not possible to connect the HV and ground lines at the same time because a large current flows. Suppose the output voltage is switched from HV to $0\ \mathrm{V}$ for example. The circuit output that is initially connected to the HV line only is disconnected to become an electrically floating state, and then it is connected to the $0\ \mathrm{V}$ line after $1\ \mu\mathrm{s}$.
%Since a large current flows, it is not possible to connect the HV part and $0\ \mathrm{V}$ part at the same time. First, the circuit output is connected to only the HV part. When switching the output voltage, the HV part is separated and the output is in an electrically floating state. After $1\ \mu\mathrm{s}$, it is connected to the $0\ \mathrm{V}$ part, and vice versa.
%

\subsubsection{HV-Switching Noise Filter}
\begin{figure}[ht]
  \centering
  \includegraphics[width=0.48\textwidth]{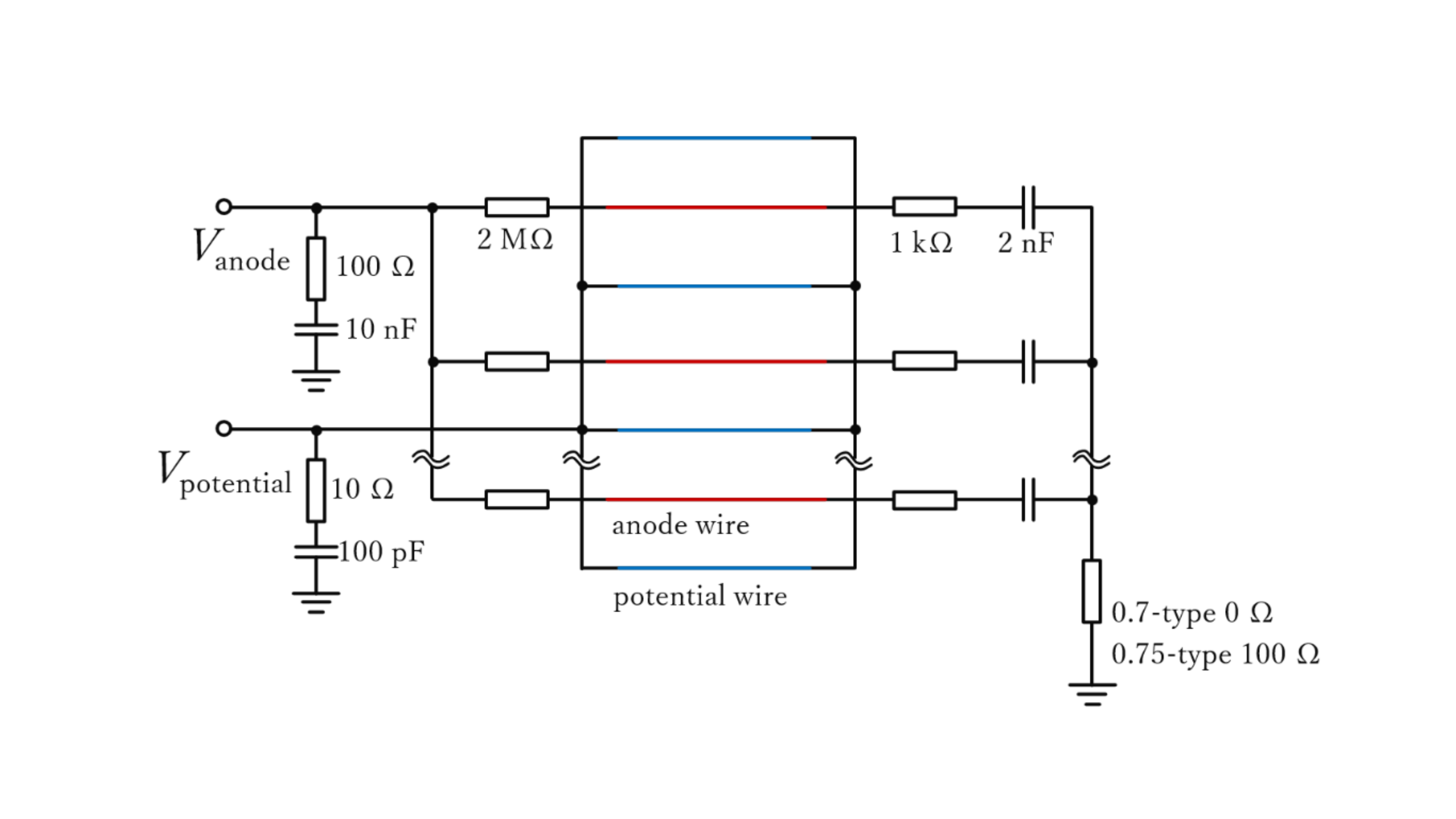}
  \caption{\label{fig:mwpccircuit}Equivalent circuit of an MWPC.}
\end{figure}

%To prevent voltage fluctuation on anode wires due to fast voltage change of potential wires, an RC filter with a $2 \mbox{ M}\Omega$  resistor and $2 \mbox{ nF}$ capacitor is attached to each anode wire.
To prevent voltage fluctuation on anode wires due to fast voltage change of potential wires, a capacitor of $2\ \mathrm{nF}$ is attached to each anode wire.
%Each anode wire is isolated from the others by $2\ \mathrm{M}\Omega$ resistor.
A $2\ \mathrm{M}\Omega$ resistor isolates each anode wire from the others.
As mentioned in~\cite{MWPC}, the capacitor value was once changed to $10 \mbox{ nF}$ to suppress electric oscillation observed in the output waveform induced by the HV switching.
In the final design, it is changed back to $2 \mbox{ nF}$ and an extra $1 \mbox{ k}\Omega$ resistor is included to reduce the total
electric current through the wires when discharge of the capacitors occurs.
% stored on the wires while suppressing the oscillation at the same time.
The snubber circuits on the voltage inputs to anode and potential wires, as shown in Fig.~\ref{fig:mwpccircuit}, are introduced for further suppression of the output oscillation.
It is also realized that the termination resistor of the anode-wire line can help reducing the oscillation amplitude, and a 100 $\Omega$ resistor is attached for the later-manufactured 0.75-type MWPCs.

\subsubsection{Output Waveform}\label{sec:Output-Waveform}
\begin{figure}[ht]
  \centering
  \includegraphics[width=0.34\textwidth, trim=30 0 -10 0]{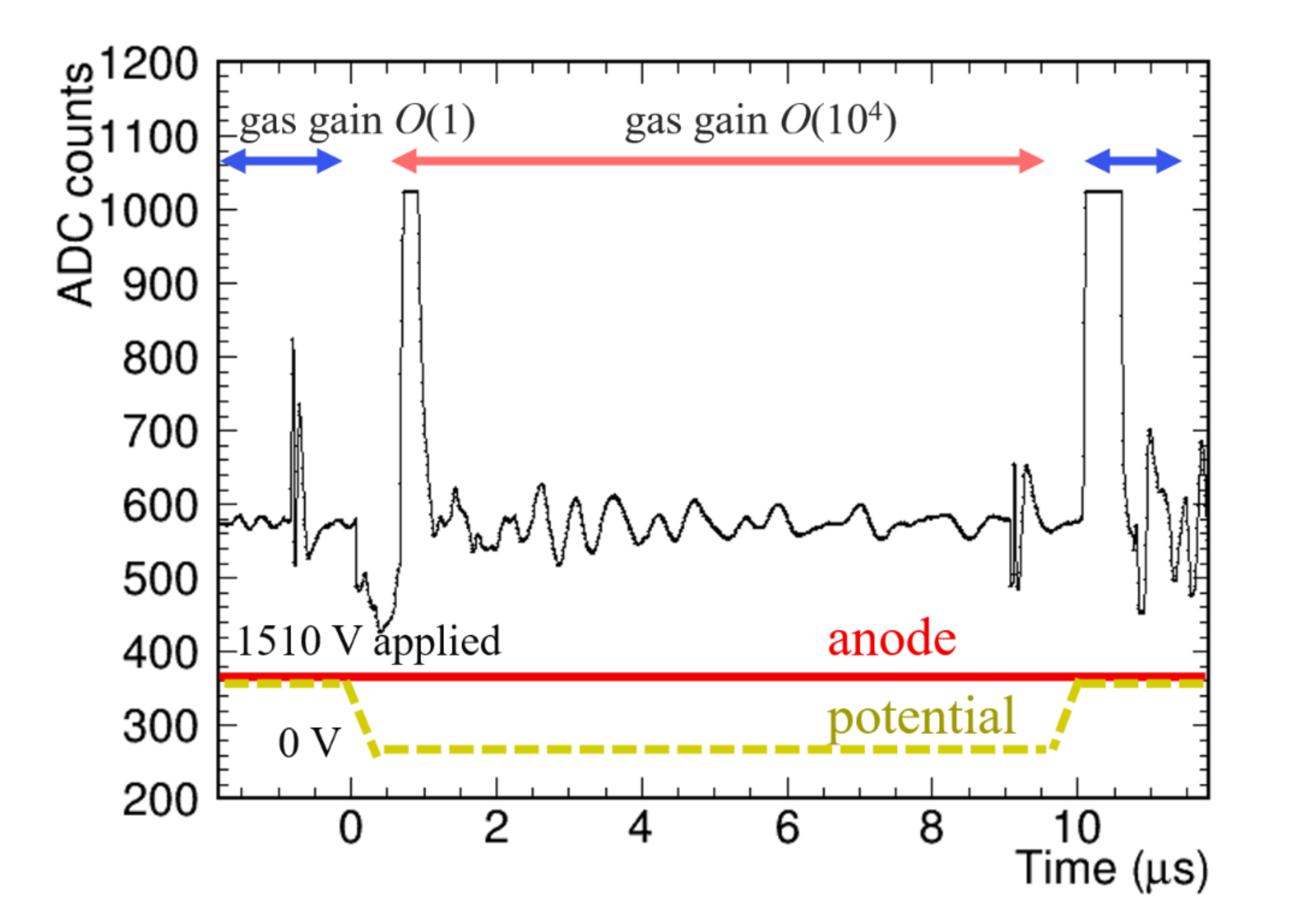}%run12985
  \caption{\label{fig:outputwaveform}Typical waveform of the detector readout due to the HV switching. The voltages applied to the anode and potential wires are also shown schematically at the bottom of the figure.}
\end{figure}

Fig.~\ref{fig:outputwaveform} shows a typical output waveform of the detector as a result of the HV switching only. 
The time when the voltage on the potential wires starts to fall is taken to be the time origin, $t=0$.
It corresponds to turning on the MWPCs.
After $t=0$, 
negative current flows into the amplifier and negative saturation occur. 
After that, due to the PZC, the waveform turns to a rapid increase to overshoot then settles down.
When the voltage is returned to the original HV, the waveform saturates positively. 
Gas multiplication occurs during the time between negative and positive saturation.
A peak followed by some response fluctuation seen at $t \simeq -1$~$\mu$s is a result of transition of the MOSFET states in the HV pulser circuit. 
A similar behavior with the opposite polarity is observed just before the positive saturation too.

The oscillation of the output after switching voltage is observed. 
It appears to be caused by the fluctuation of the circuit for HV switching. 
It is still possible to find a signal by subtracting a template waveform consisting of the most frequent amplitude obtained from a few hundred waveforms because the shape of the oscillation is rather stable and unchanged.

%--------------------------------------------------
\subsection{Operational Conditions}
\subsubsection{Discharge Test}\label{sec:discharge}
Because the wire pitch between the anode and potential wires is rather small, it is important to understand discharge voltages for stable operation of the MWPCs.
In fact, it was observed for an MWPC that a significant number of wires were cut simultaneously when discharge occurred.

\begin{figure}[tbp]
  \centering
  \includegraphics[width=0.4\textwidth]{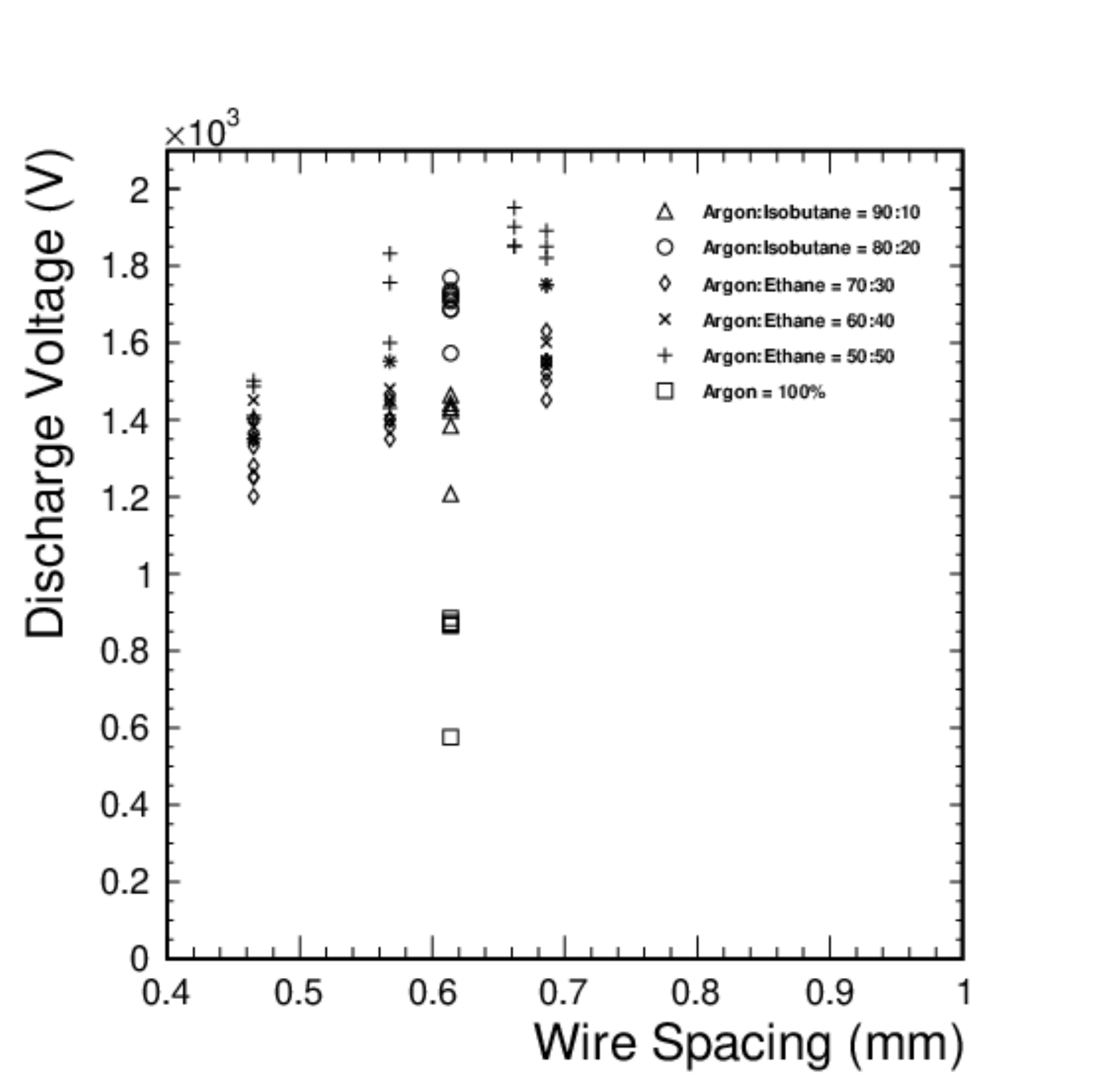}
  \caption{\label{fig:dischargevoltage}Discharge voltages for several different gas mixtures at atmospheric pressure. Variation of the data points of the same marker type for a given wire spacing represents the reproducibility of the measurement.}
\end{figure}

At the R\&D stage, discharge voltages were measured for the nominal wire pitches of 0.5 mm, 0.6 mm, and 0.7 mm.
Fig.~\ref{fig:dischargevoltage} shows the result obtained by using several different gas mixtures at atmospheric pressure.
This measurement was performed using an anode and potential wires tensed on a glass epoxy board in a small chamber.
The values of wire spacing were actual measurements by a microscope.
We set the potential wire at $0\ \mathrm{V}$, while we increased the voltage to the anode wire at a ramping speed of $1\ \mathrm{V/s}$~\cite{MWPC, Takezaki}.

According to the Paschen’s law~\cite{Paschen}, the discharge voltage is approximately proportional to the distance between the electrodes if the distance is in a range between $0.1$ and $1\ \mathrm{mm}$.
Using this law, the discharge voltage for a distance of 0.7 mm is calculated for each measurement shown in Fig.~\ref{fig:dischargevoltage}, and the lowest possible voltage is derived as conservative estimation. It is found to be 1760 V
%For a distance of $0.7\ \mathrm{mm}$, the lowest discharge voltage is therefore \Red{$1760$} V 
for argon/ethane $=$ 50\%/50\%, $1790\  \mathrm{V}$ for argon/isobutane $=$ 80\%/20\%, 
$1380\ \mathrm{V}$ for argon/isobutane $=$ 90\%/10\%, and $660\ \mathrm{V}$ for argon $=$ 100\%. 
%
%\newpage

%\begin{spacing}{1.2}
%
%\mbox{}
%\vspace{20pt}
%
\subsubsection{Gas Gain}
\begin{figure}[tbp]
  \centering
  \includegraphics[width=0.4\textwidth]{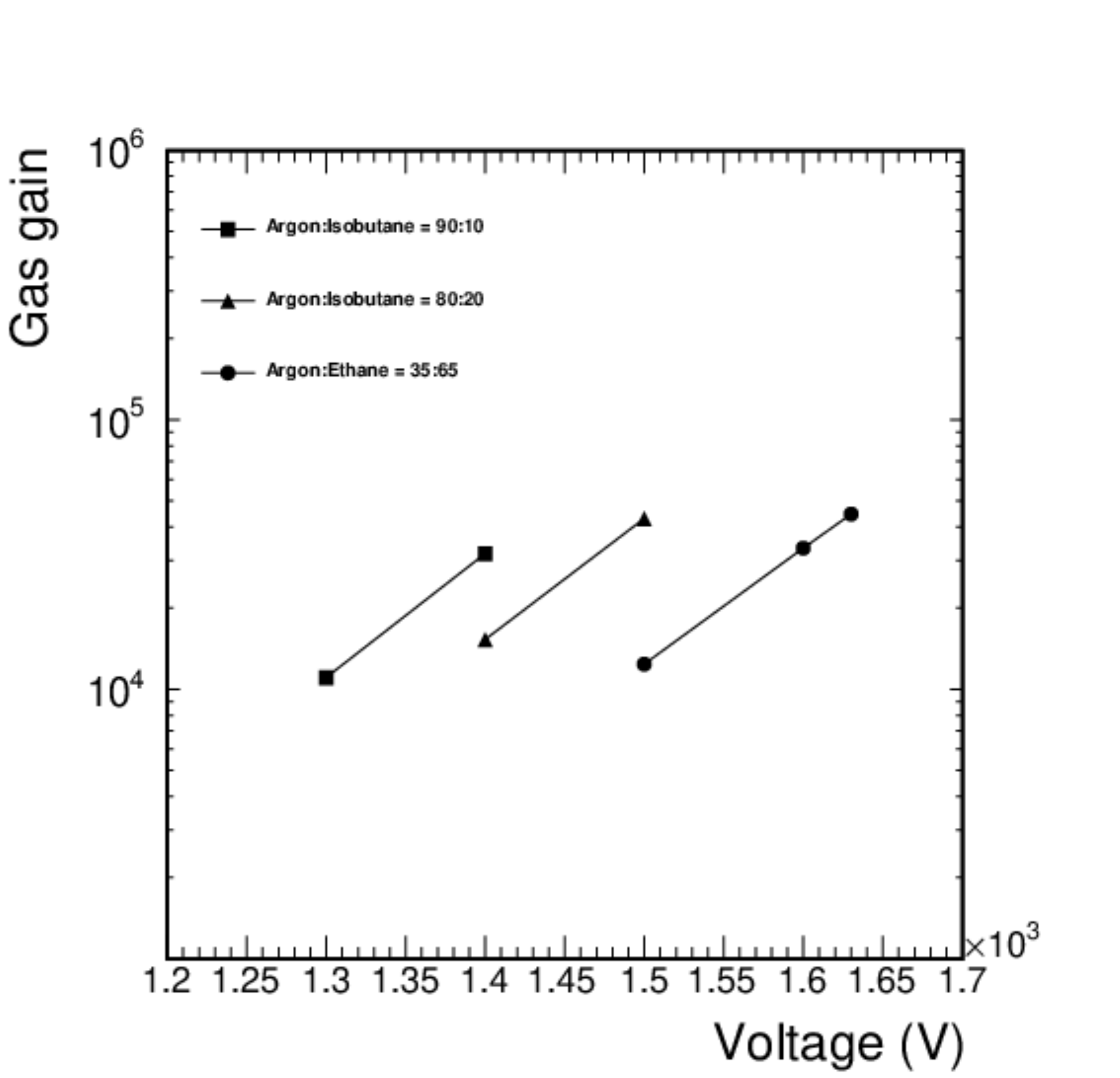}
  \includegraphics[width=0.4\textwidth]{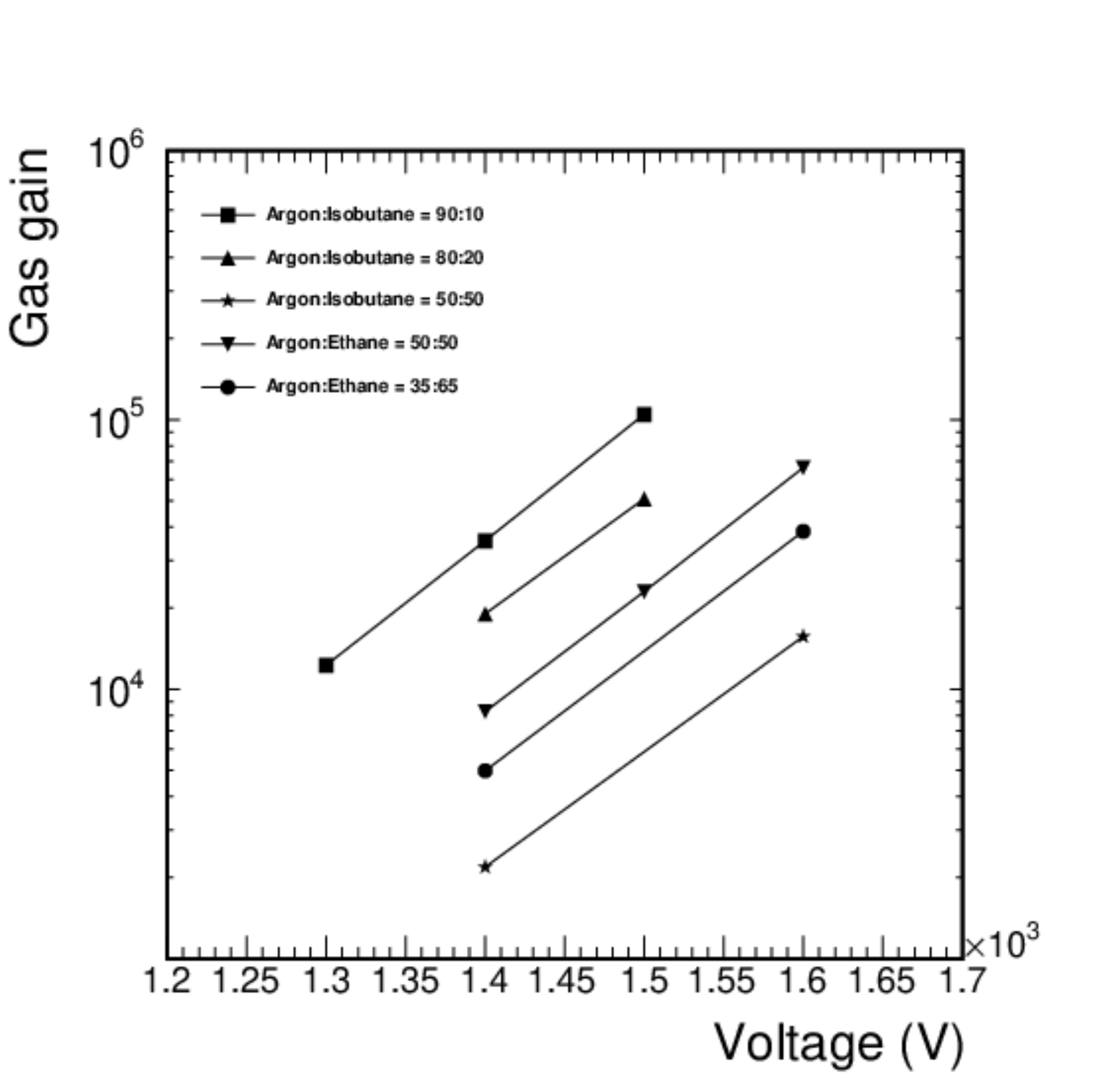}
  \caption{\label{fig:gasgain}Simulated gain of gas multiplication as a function of applied voltage to the anode wires with the potential wires at 0 V for wire spacings of 0.75 mm (top) and 0.7 mm (bottom).}
\end{figure}

When a charged particle is incident on the MWPC, electron-ion pairs are created. The mean number of electron-ion pairs created between the two cathode planes with a gap of 6 mm is approximately 62 pairs for argon/ethane = 50\%/50\%.
Gas multiplication occurs if a strong electric field exists around the anode wire. Fig.~\ref{fig:gasgain} shows the mean gain of gas multiplication as a function of applied voltage to anode wires estimated by Garfield++~\cite{3} for several cases of gas mixtures, where the voltage of the potential wire is set to 0 V. 
In this simulation, electrons are randomly placed at a distance of $150\ \mu\mathrm{m}$ from the center of the anode wires in a chamber in which the anode and potential wires are tensed alternately with an interval of $0.7\ \mathrm{mm}$ or $0.75\ \mathrm{mm}$, 
and the number of ions created after avalanche multiplication is counted.

For a gas gain of $5 \times 10^{4}$ with a wire spacing of $0.7\ \mathrm{mm}$, the required voltage is $1580\ \mathrm{V}$ with argon/ethane $=$ 50\%/50\%, $1500\ \mathrm{V}$ with argon/isobutane $=$ 80\%/20\%, and $1440\ \mathrm{V}$ with argon/isobutane $=$ 90\%/10\%. 
By looking at the discharge voltages we discussed in Section~\ref{sec:discharge}, the margin voltages to discharge are $180$ V, $290\ \mathrm{V}$, and $-60\ \mathrm{V}$ (unstable due to discharge), respectively.

\begin{figure}
\begin{overpic}[width=0.43\textwidth, trim=5 0 10 0, clip]{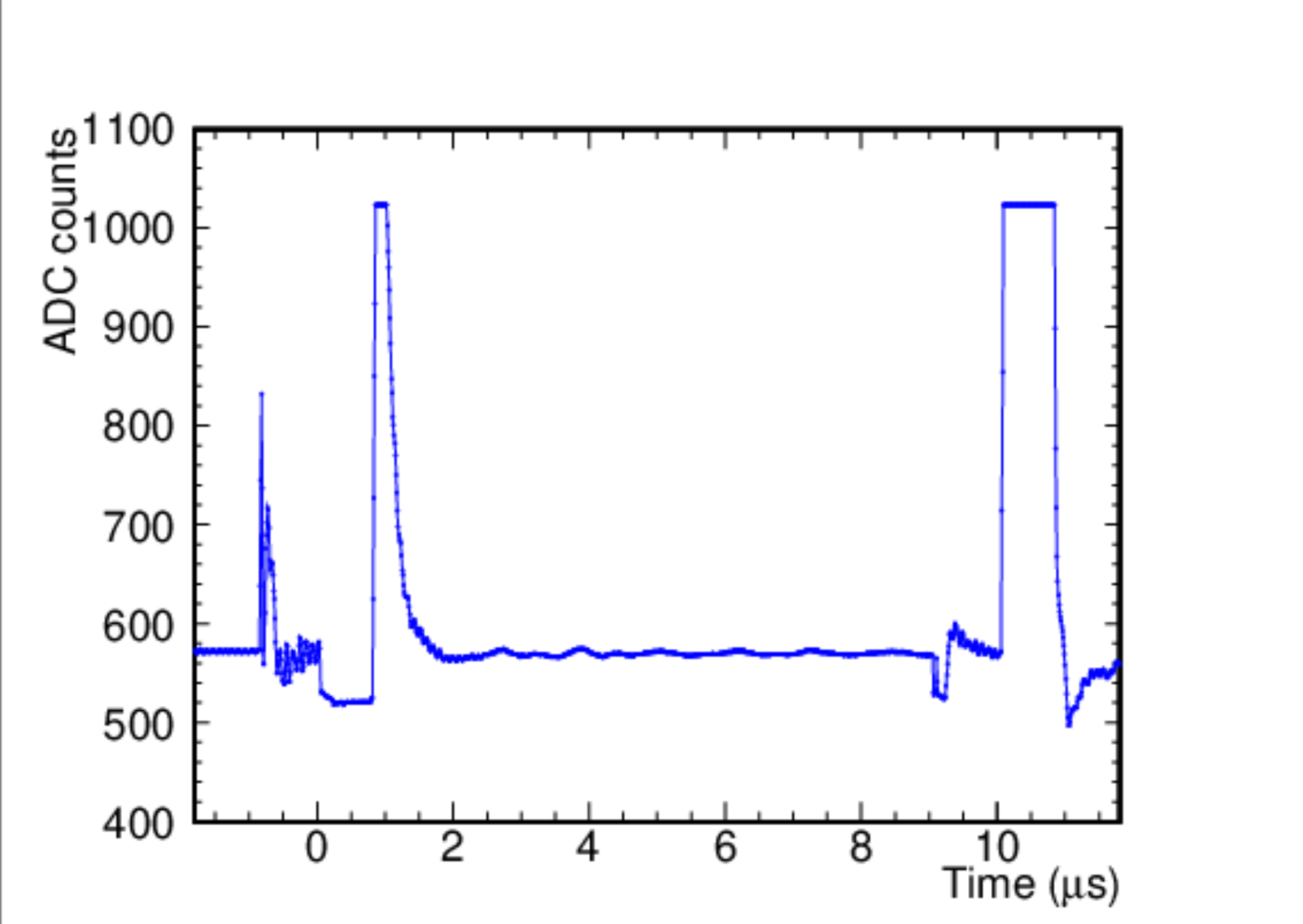}
  \put(40,12){1000 V applied}
\end{overpic}
\begin{overpic}[width=0.43\textwidth, trim=5 0 10 0, clip]{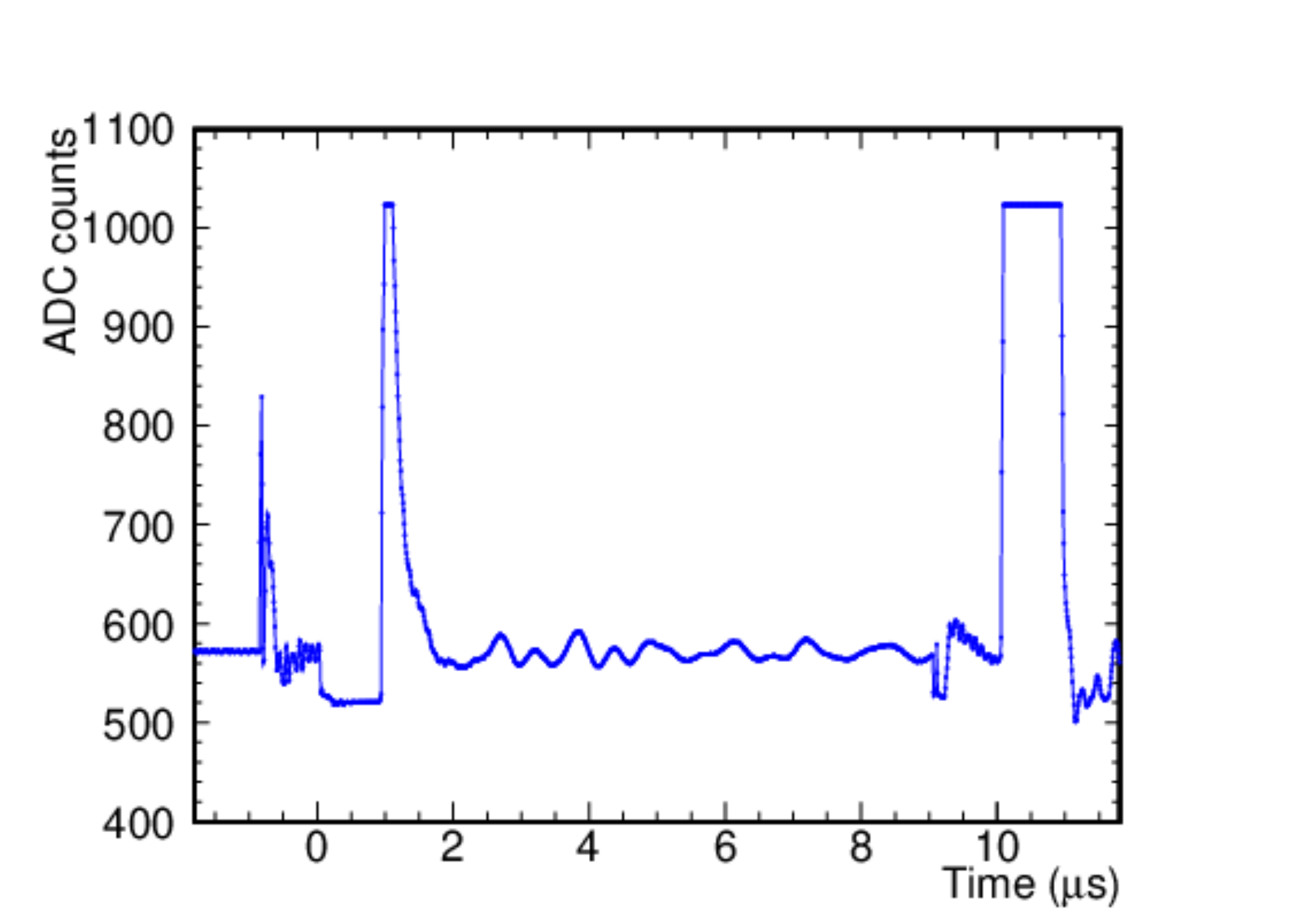}
  \put(40,12){1300 V applied}
\end{overpic}
\begin{overpic}[width=0.43\textwidth, trim=5 0 10 0, clip]{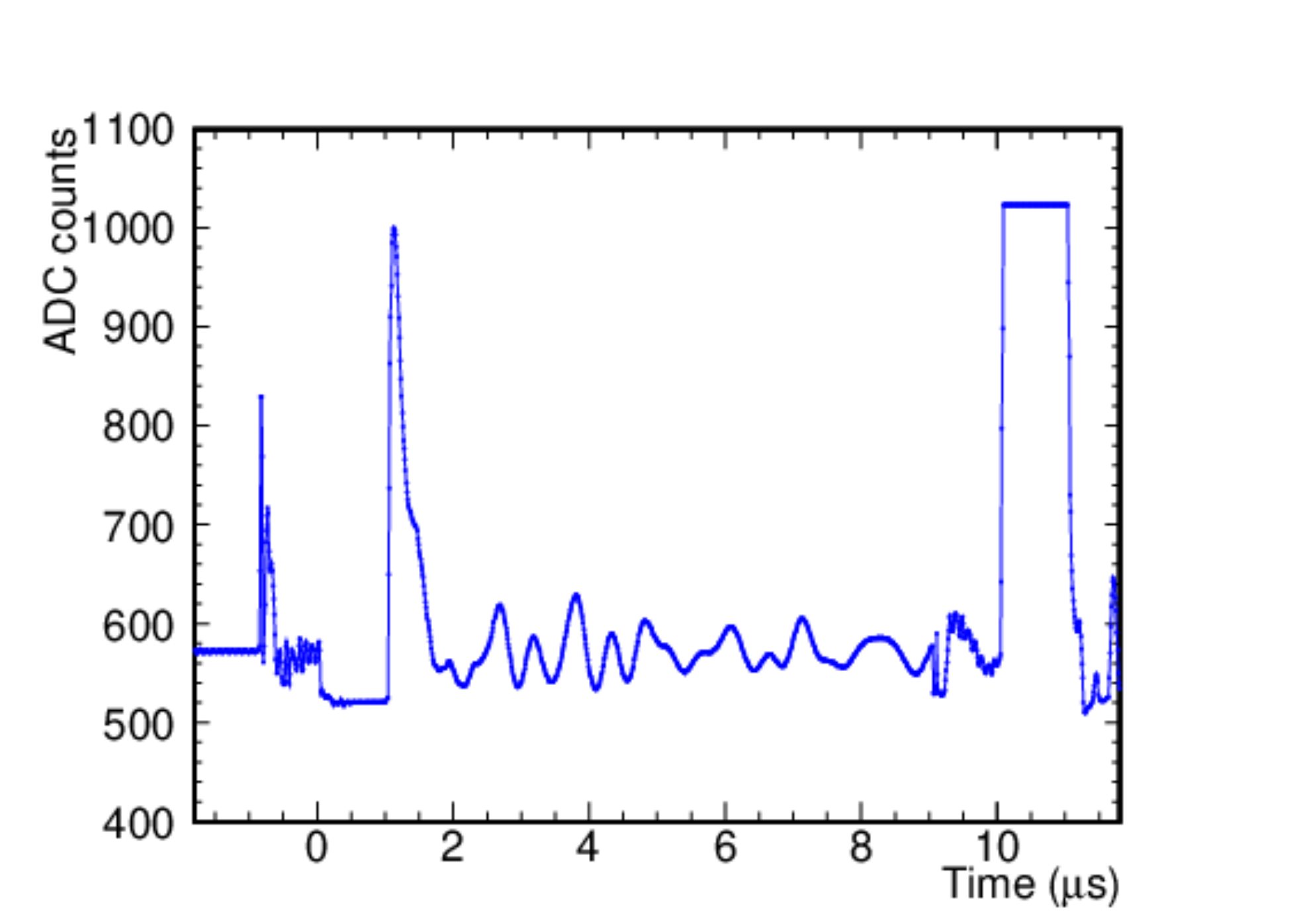}
  \put(40,12){1650 V applied}
\end{overpic}
\caption{\label{fig:wfandHV}Typical waveforms for different applied HV values.}
\end{figure}

The amplitude of oscillation in the output waveform becomes larger as the applied voltage increases, as shown in Fig.~\ref{fig:wfandHV}. 
%
%
%\begin{figure}[H]
%\begin{figure}
%  \centering
%  \includegraphics[width=0.4\textwidth]{images/wfandHV2.png}%run 5749 to 5753 FI03 ch23
%  \caption{\label{fig:wfandHV}Typical waveforms for different applied HV values.}
%\end{figure}
%
\noindent To avoid negative saturation of the waveform as a result of the oscillation, the HVs are set as low as possible to keep sufficient gain and ensure stable operation. Thus, argon/isobutane $=$ 80\%/20\% is adopted as the base gas mixture.
%
%\end{spacing}
%
%\clearpage

%==================================================
\section{Hit Finding in Waveform}
%==================================================
As already described in the previous sections, the baseline of the read-out signal from this MWPC is not flat due to the induced noise of HV-switching.  It is impossible to use a simple discriminator and time-to-digital converter to extract hit information. To solve this problem, the waveform of the read-out signals is recorded using a 100-MHz Fast-ADC, and the bumpy baselines are subtracted in the offline analysis.  The computer algorithm to extract the hit information from the waveform is as follows:
\begin{enumerate}
    \item Subtraction of the template waveform.
    
    The template waveform (solid line in Fig.~\ref{fig:subt}) is constructed by identifying the most frequent ADC count at each FADC sample point obtained from several hundred waveforms.
    %
    %
%    \begin{figure}[H]
    \begin{figure}
%\begin{minipage}{0.45\hsize}
  \centering
  \includegraphics[width=0.45\textwidth]{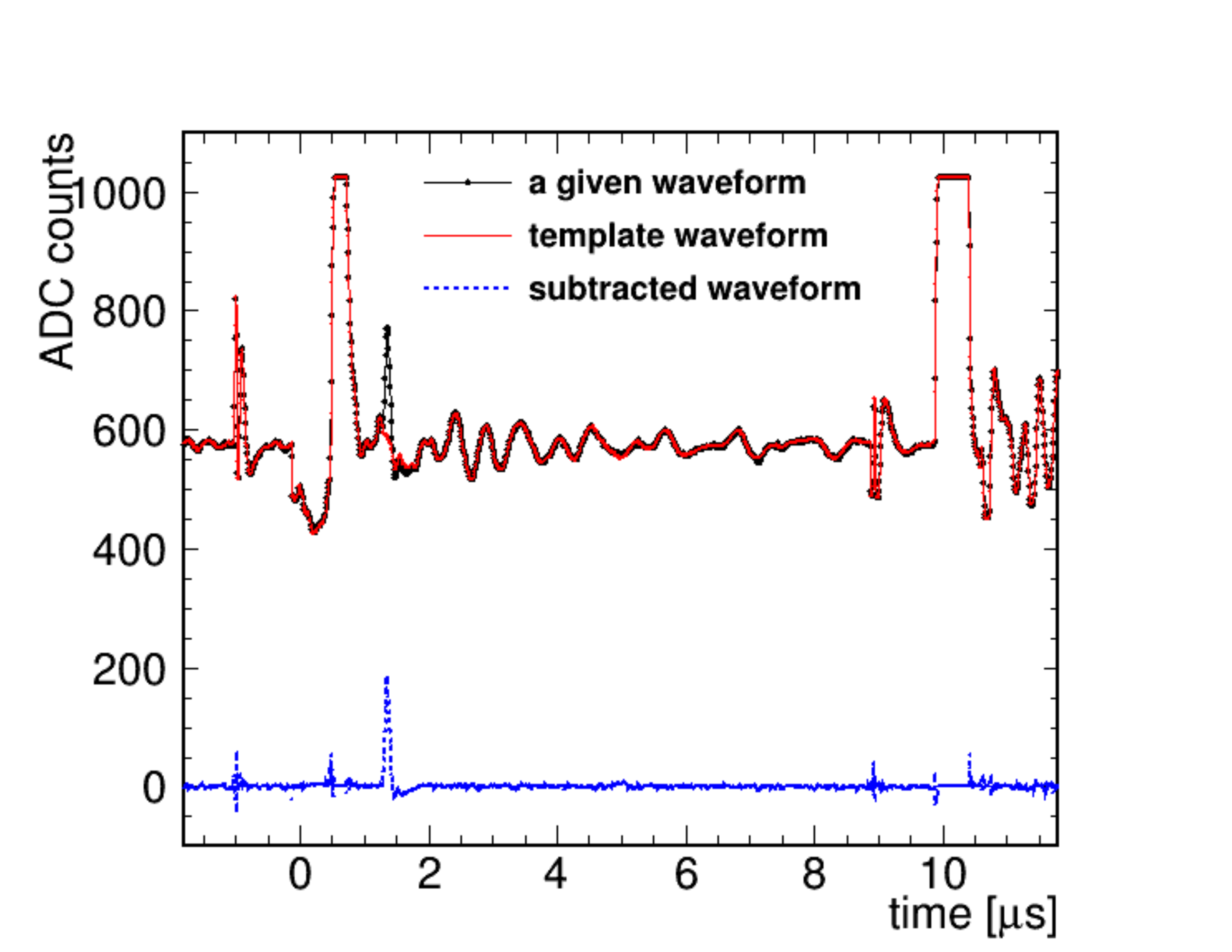}%run12985
  \caption{\label{fig:subt}Template waveform (solid line) consisting of the most frequent ADC counts, a waveform in a certain trigger (line with dots), and the subtracted waveform (dashed line).}
\end{figure}
    The dashed line in Fig.~\ref{fig:subt} is the waveform after subtracting the template from a given waveform, shown as the solid line with dots.
    A peak at around $1.5$ $\mu$s in the dashed line corresponds to the signal.
    Small fluctuations seen in the time region within $\pm 1$ $\mu$s and after 8.5 $\mu$s arise from imperfect subtraction of the template waveform due to FADC jitters.
    Concerning the HV switching and the noise induced by that, 
    please refer to Fig.~\ref{fig:outputwaveform} (\S\ref{sec:Output-Waveform}).
    Signal and pedestal pulse heights are shown in Fig.~\ref{fig:SigPed}, where the signal pulse height is a response sum of three channels around the strip with the largest signal, and the pedestal values are calculated for two time-regions, before and after the HV switching.
    Two pedestal distributions are almost the same and the subtraction of the template waveform is confirmed to be effective to extract signals.
    \begin{figure}
    \centering
    \includegraphics[width=0.34\textwidth, trim=0 0 45 0]{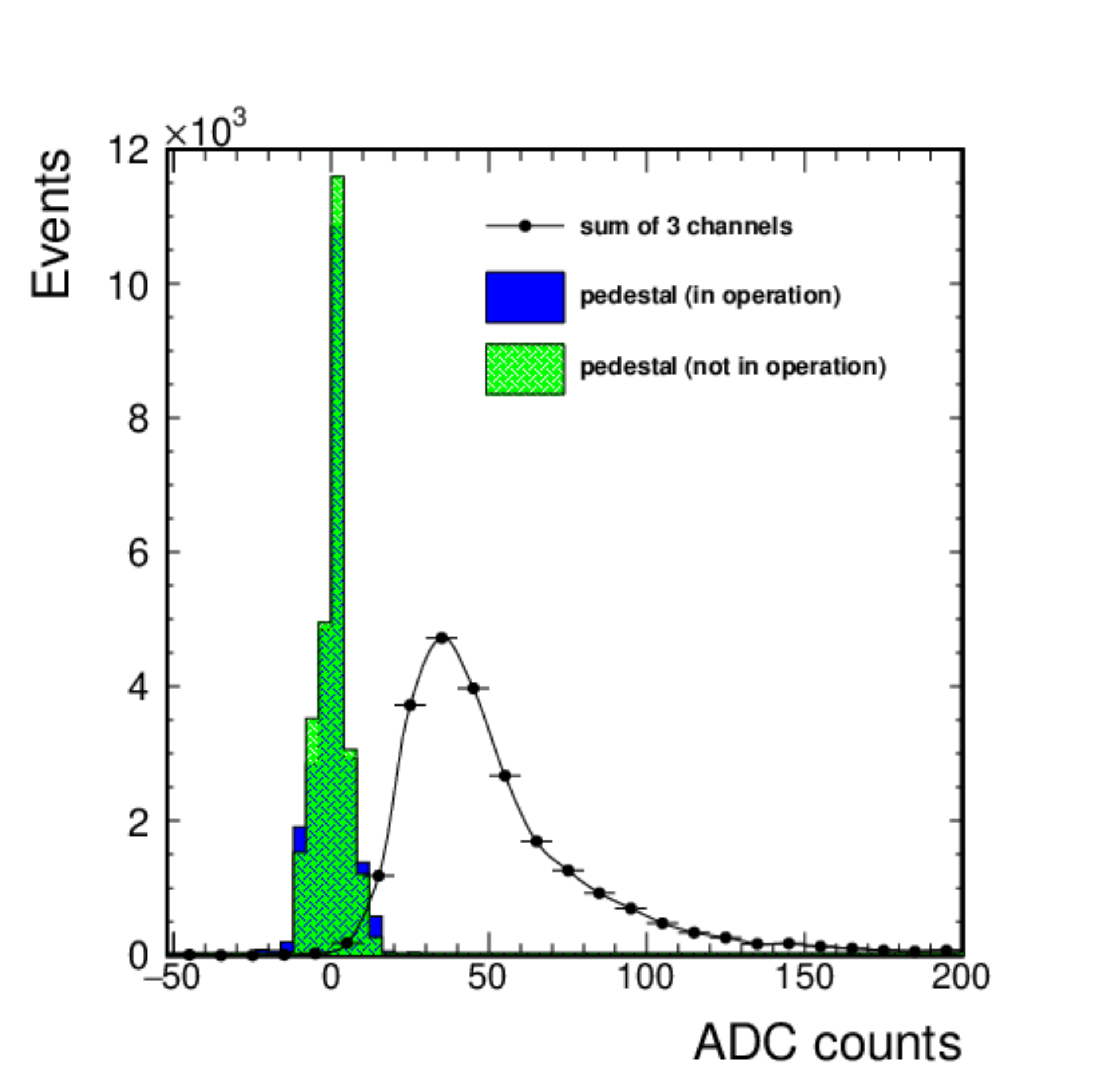}
    \caption{Distribution of signal pulse heights and pedestal values.}
    \label{fig:SigPed}
    \end{figure}

    \item Cluster construction.
    
    For each strip and each FADC sample point, a sum of ADC counts over a certain range of the strip and the time direction is calculated, which we call a ``cluster''.
    As shown in Fig.~\ref{fig:chargedist}, using five $x$-strips around a given strip, the ADC counts of the three center channels are summed up while subtracting the average ADC count calculated from the outer two channels as the common noise level. 
    %
    %
%    \begin{figure}[H]
    \begin{figure}
%\begin{minipage}{0.45\hsize}
  \centering
  \includegraphics[width=0.35\textwidth, trim=0 0 0 10]{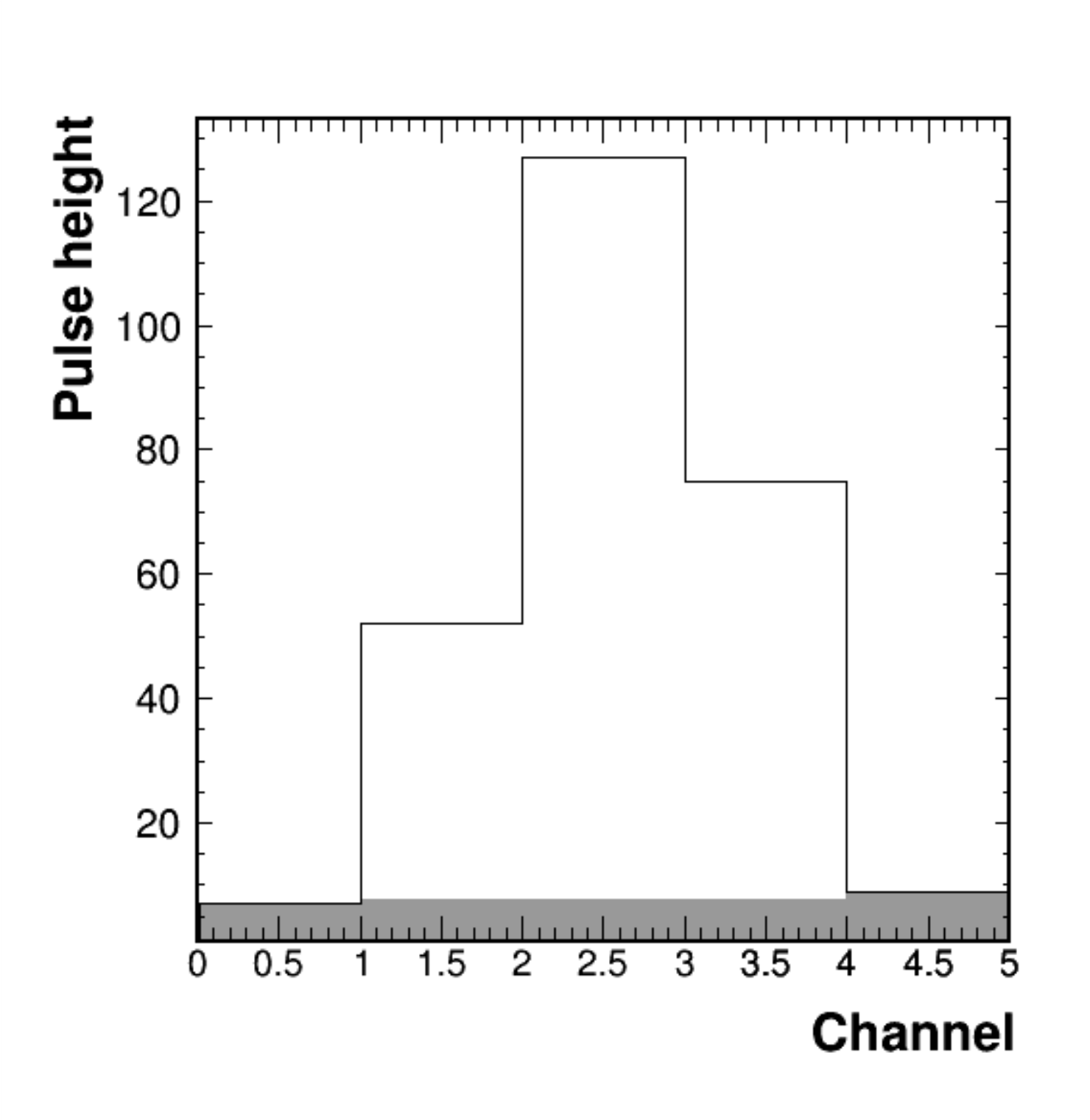}
  \caption{\label{fig:chargedist}Example of pulse heights in five cathode strips with a signal.}
%\end{minipage}
\end{figure}
    Then, these 3-strip sums are added over ten sample points in the time direction starting from the sample point under consideration because the FWHM of signal responses is approximately 100 ns, independent of pulse heights, as shown in Fig.~\ref{fig:wfs}.
    %
    %
%    \begin{figure}[H]
    \begin{figure}
\begin{center}
%\begin{minipage}{0.45\hsize}
\includegraphics[width=0.32\textwidth, trim=0 0 20 0]{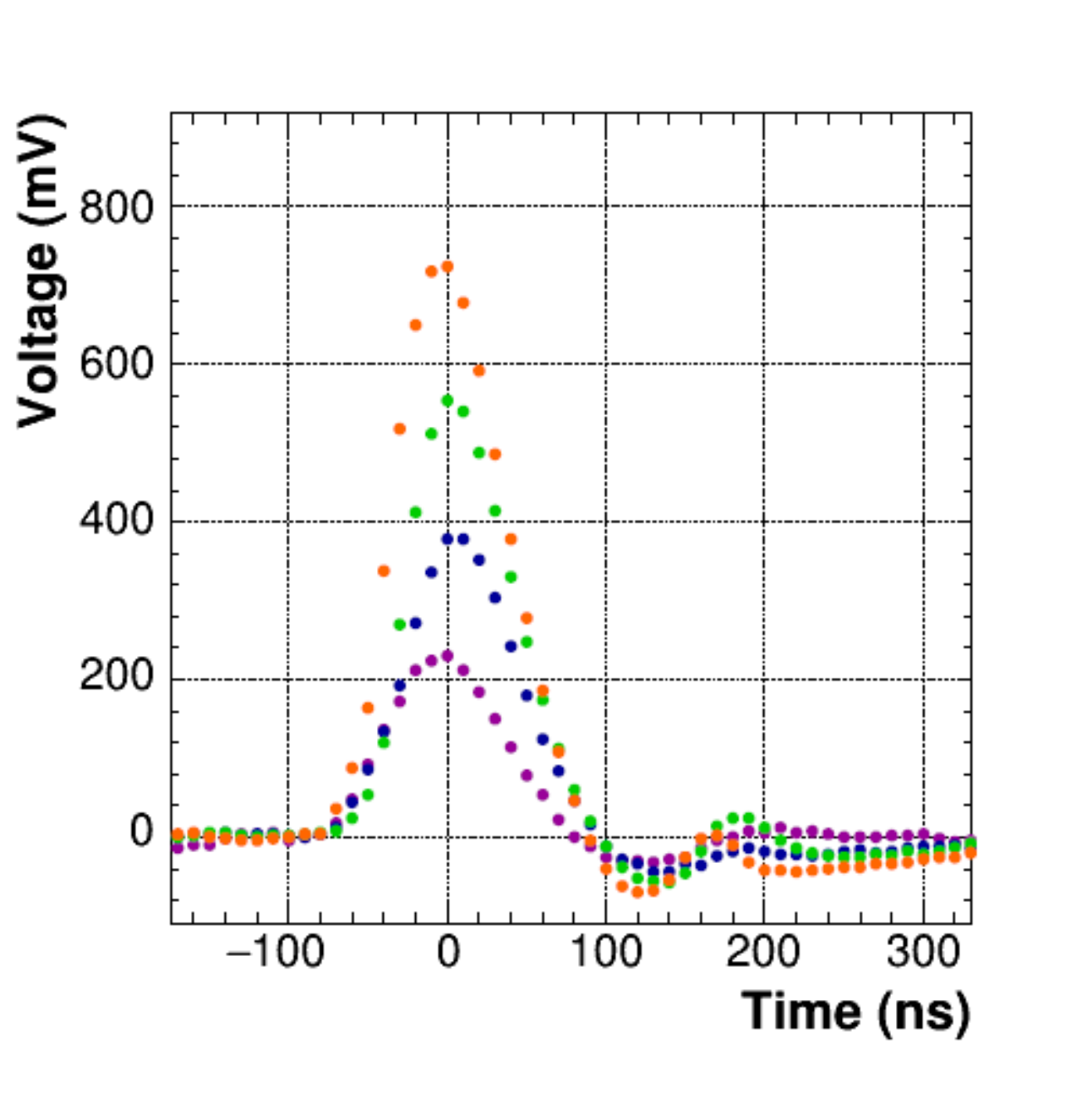}
\caption{\label{fig:wfs}Signals with various pulse heights. Waveforms from different events are superimposed.}
%\end{minipage}
\end{center}
\end{figure}

    \item Hit finding and position reconstruction.
    
    If one cluster larger than a threshold is found, the local maximum around the cluster within a region of $\pm 2$ strips and $\pm 2$ sample points is identified.
    The local maximum cluster is accepted as a hit if three consecutive clusters in the time direction around the local maximum cluster are larger than the threshold.
    The hit position for a strip channel $i$ is calculated by the center of mass method using the three strips of the cluster as $\sum_{j=i-1}^{i+1} (j \cdot Q_{j}) / \sum_{j=i-1}^{i+1} Q_{j}$, where $Q_{j}$ and $j$ are the strip ADC count (summed over 10 FADC sample points) and strip channel number, respectively.
\end{enumerate}
%
%\end{minipage}
%\mbox{}
%
%
%==================================================
\section{Test and Performance Evaluation}
\begin{figure}[htbp]
    \centering
    \includegraphics[width=0.4\textwidth]{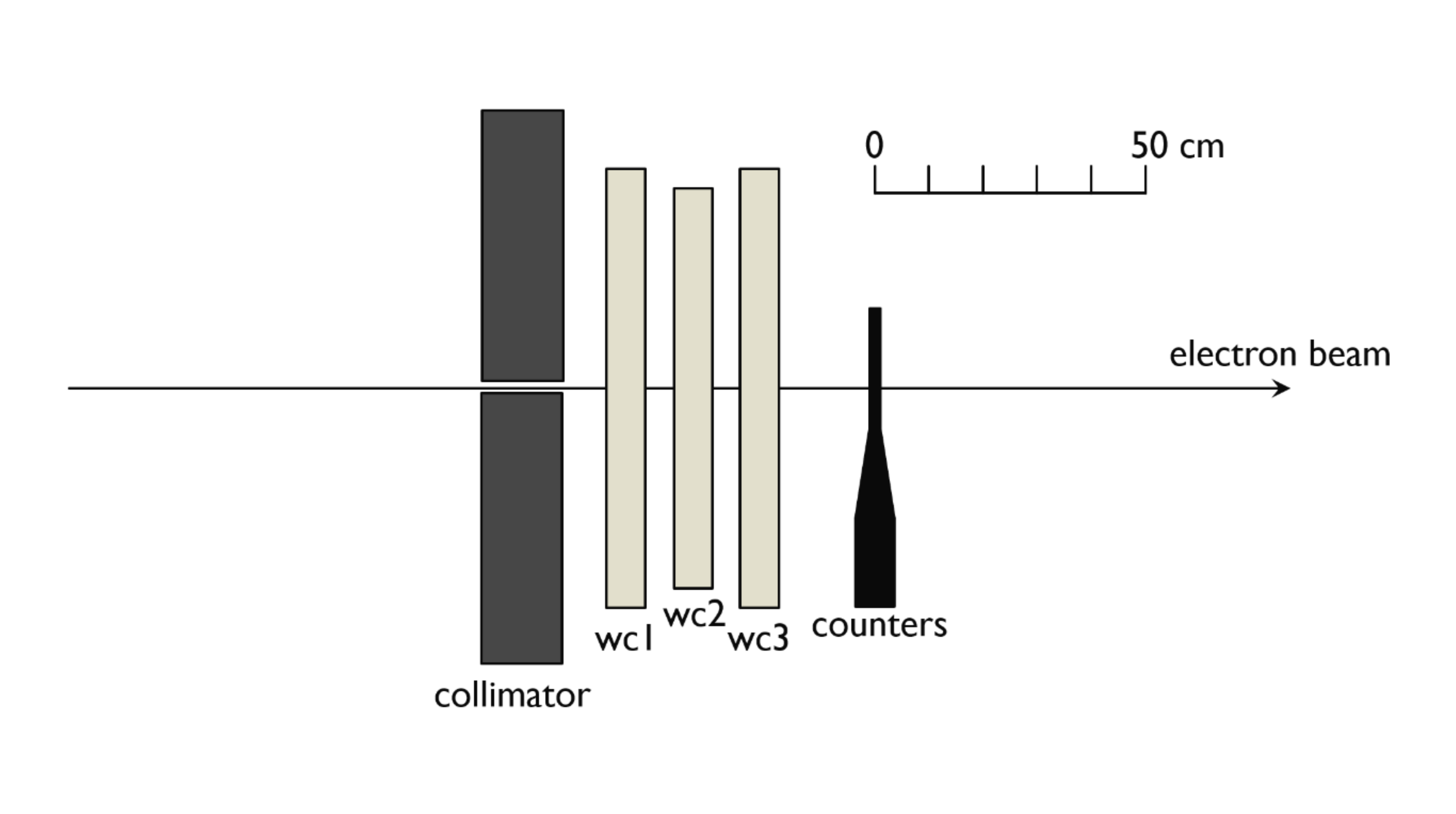}\\
    \includegraphics[width=0.4\textwidth]{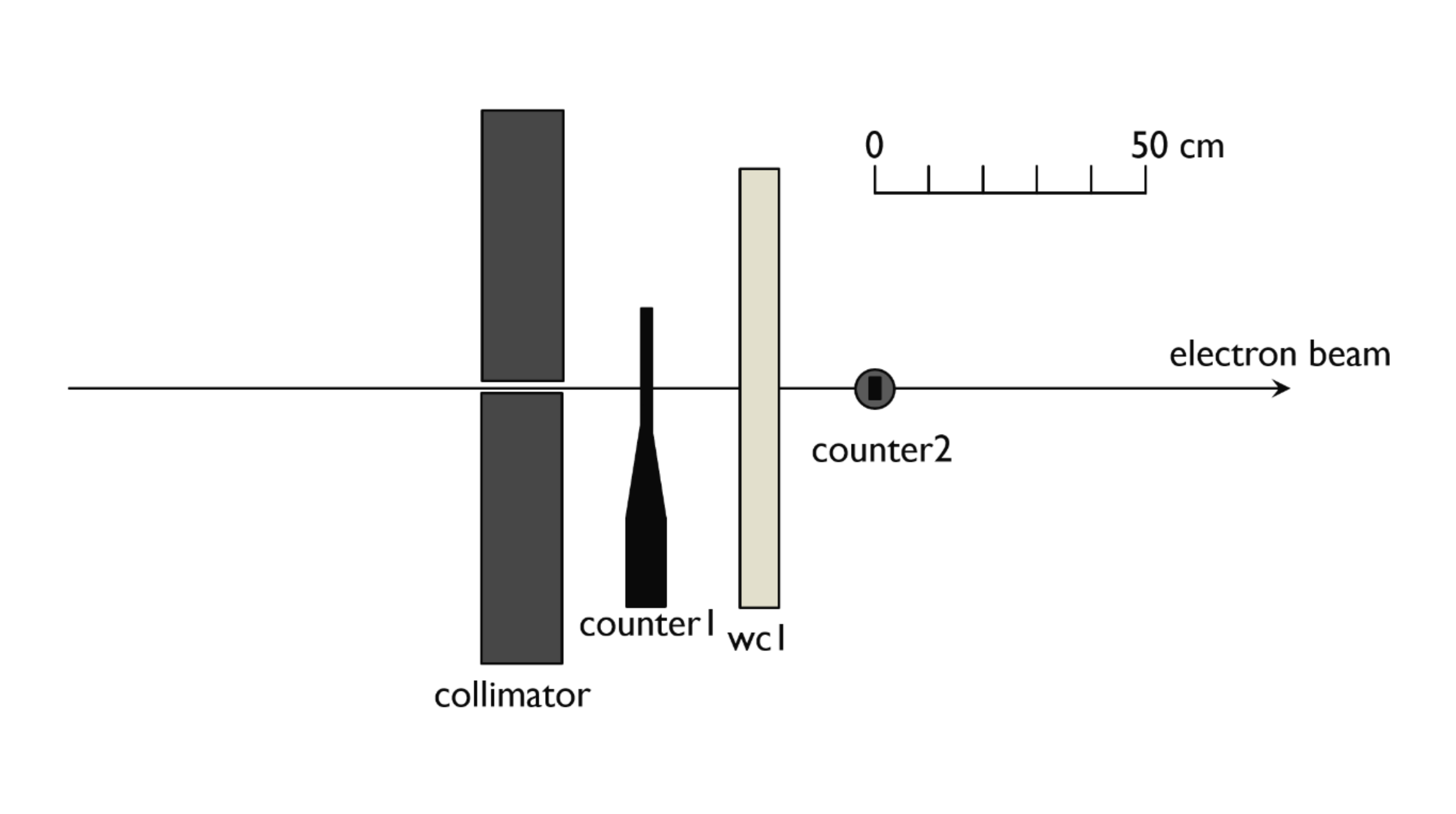}
    \caption{\label{fig:setupkurns}Top view of the experimental setups for the measurement of pulse height and position resolution (above) and the measurement of hit efficiency (below). Electron beam is collimated with lead blocks. WC1, WC2, and WC3 are the production-type MWPCs. WC1 and WC3 are 0.75-type and WC2 is 0.7-type. Scintillation plastic counters are placed to measure the beam intensity.}
    \end{figure}

%==================================================
The performance of MWPCs was evaluated by using electron beams of the Linac at Kyoto University Institute for Integrated Radiation and Nuclear Science.
Fig.~\ref{fig:setupkurns} shows the experimental setup.
The electron beam was collimated to 10 mm wide and 18 mm high with lead and iron blocks.
The repetition rate of the pulsed beam was set to 25 Hz, which was the same as the RCS in the real experiment.
At the beam exit, the MWPCs are placed with scintillation plastic counters for counting the number of electrons. 
The duration of the beam pulse was 4 $\mu$s and the number of electrons was maintained to be approximately a few per pulse.
For clarification, we note that beam particles to simulate large number of prompt charged particles did not irradiate MWPCs in this basic performance evaluation.
%In this performance evaluation, no prompt charged particle was applied. 
The default beam energy for our tests was 16 MeV with a FWHM of 1.2 MeV, but in order to investigate an energy dependence of the hit position resolution mainly due to multiple scattering effects, the electron beam with 30 MeV was also used~\cite{Linac}.
%
%--------------------------------------------------
\subsection{Pulse Height}\label{sect4.1}
\begin{figure}[htbp]
    \centering
%\begin{minipage}{0.45\hsize}
    \includegraphics[width=0.4\textwidth]{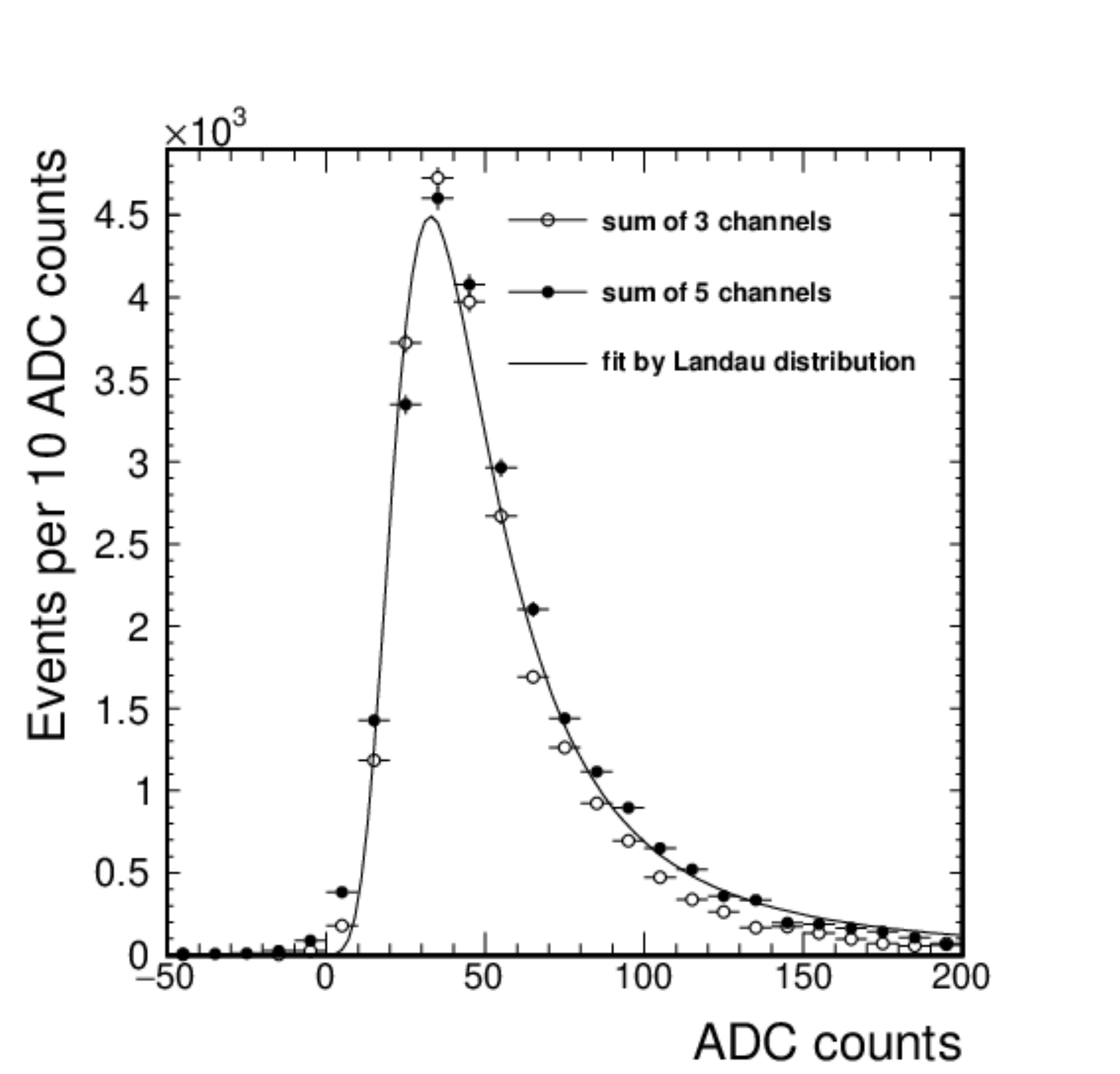}
    \caption{\label{fig:3pkand5pk}Pulse height distribution in 1 ADC count $=2.0\ \mathrm{mV}$. Open circles correspond to the sum of ADC counts of three channels, the cathode strip with the highest pulse height and two adjacent strips, and black points show the sum of five channels.}
%\end{minipage}
    \end{figure}

Fig.~\ref{fig:3pkand5pk} shows the pulse height distribution.
The distribution denoted by open circles corresponds to the sum of the ADC counts of three channels, the cathode strips with highest pulse height and two adjacent strips, while the distribution represented by black circle points shows the sum of five channels.
%The 3pk data in black points is the sum of the adc counts of three channels, the cathode strips with the highest pulse height and those on both sides, while the 5pk data in blue points is the sum of five channels. 
These are well represented by the Landau distribution. Because there is not much difference in the shape of distribution between three and five channels, it can be said that the avalanche charge created by an incident particle is within the size of $3\ \mathrm{mm}\ \times\ $3 cathode strips. The average avalanche charge is $\simeq17\ \mathrm{fC}$, where readout amplifiers with a gain of $6.9\ \mathrm{V/pC}$ and FADCs with an ADC count of $2.0\ \mathrm{mV}$ are used.

%--------------------------------------------------
\subsection{Hit Efficiency}
\begin{figure}[htbp]
    \centering
    \includegraphics[width=0.4\textwidth]{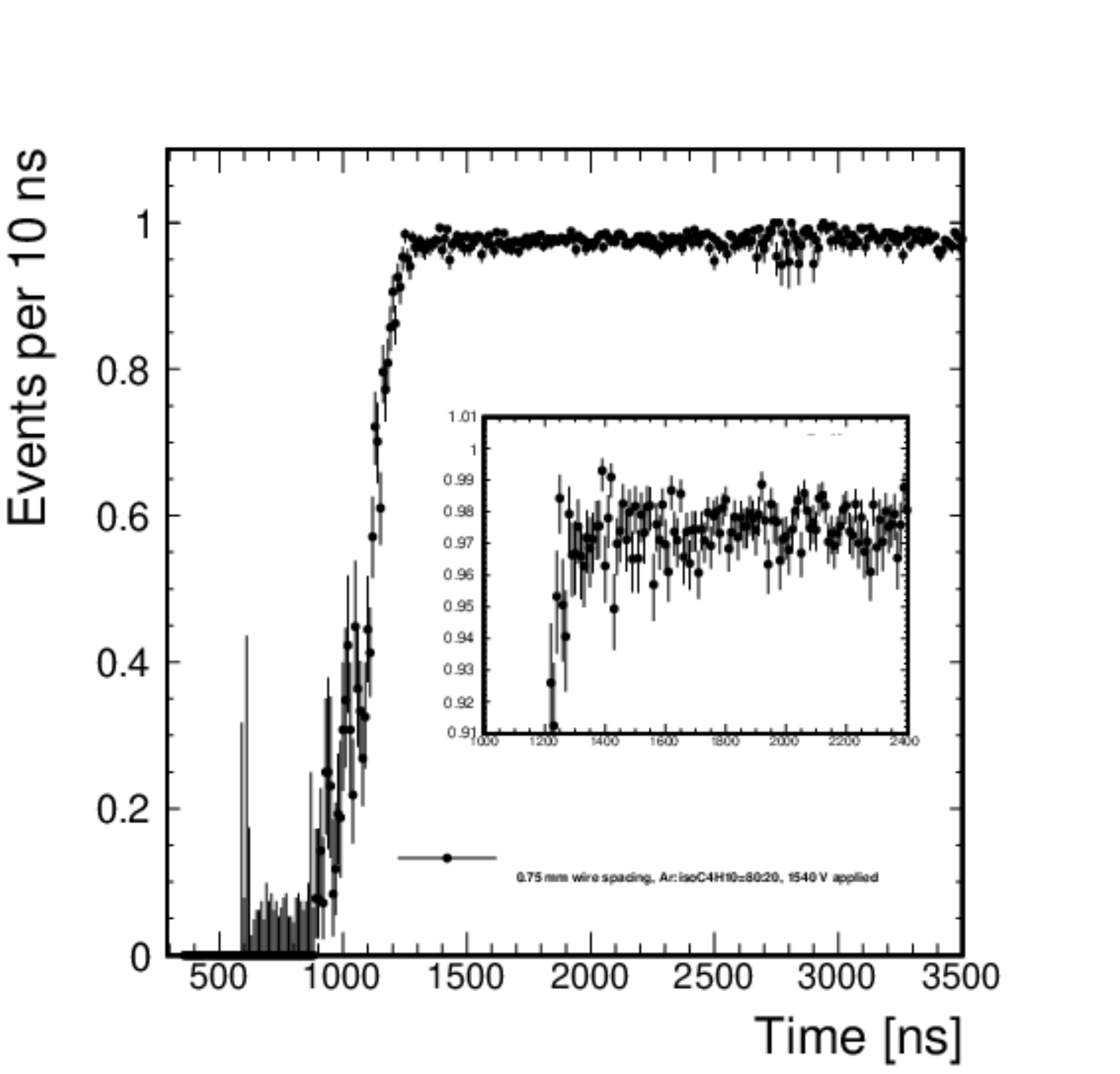}
    \caption{\label{fig:hitefficiency}Single hit efficiency of the MWPC as a function of time after the voltage on the potential wires starts to fall.}
%
%note10 p148
%
\end{figure}

The hit finding efficiency is estimated by looking at the fraction of coincidences between the two counters with a hit found in the MWPC.
More concretely, the coincidence hits within 10 ns detected by the two counters are considered to correspond to electrons penetrating the MWPC.
For a given coincidence, strip responses of the MWPC corresponding to the beam position are examined and the hit that is found by the algorithm and is matching with the coincidence time within 100 ns is defined to be the signal successfully identified by the MWPC.
Fig.~\ref{fig:hitefficiency} shows the efficiency as a function of time 
after turning on the MWPC.
The MWPC with a wire spacing of $0.75\ \mathrm{mm}$ is filled with a mixed gas containing argon/isobutane $=$ 80\%/20\%.
A DC of $1540\ \mathrm{V}$ and switching voltage as shown in Fig.~\ref{fig:methodtoapplyHV} 
(middle) with a width of $10\ \mu\mathrm{s}$ are applied to the anode and potential wires, respectively. 
The efficiency for a single electron becomes approximately 98\% in 1.4 $\mu$s after turning on the operation of the MWPC.
This recovery time corresponds to a 50\% efficiency for signals from muons of muonic carbon-atoms with a lifetime $\simeq 2.0$ $\mu$s.
The dead time will be shortened by optimizing parameters such as the resistor value at the output line of the HV pulser for the time constant of the HV change, the resistor value of the PZC for the time duration of the positivive saturation, both having effects of a few handreds nanoseconds, using actual beams in the real experiment.

\subsection{MWPC Position Resolution and DeeMe experiment}
\begin{figure}[htbp]
\begin{minipage}{0.49\hsize}
\begin{center}
\includegraphics[width=0.9\textwidth]{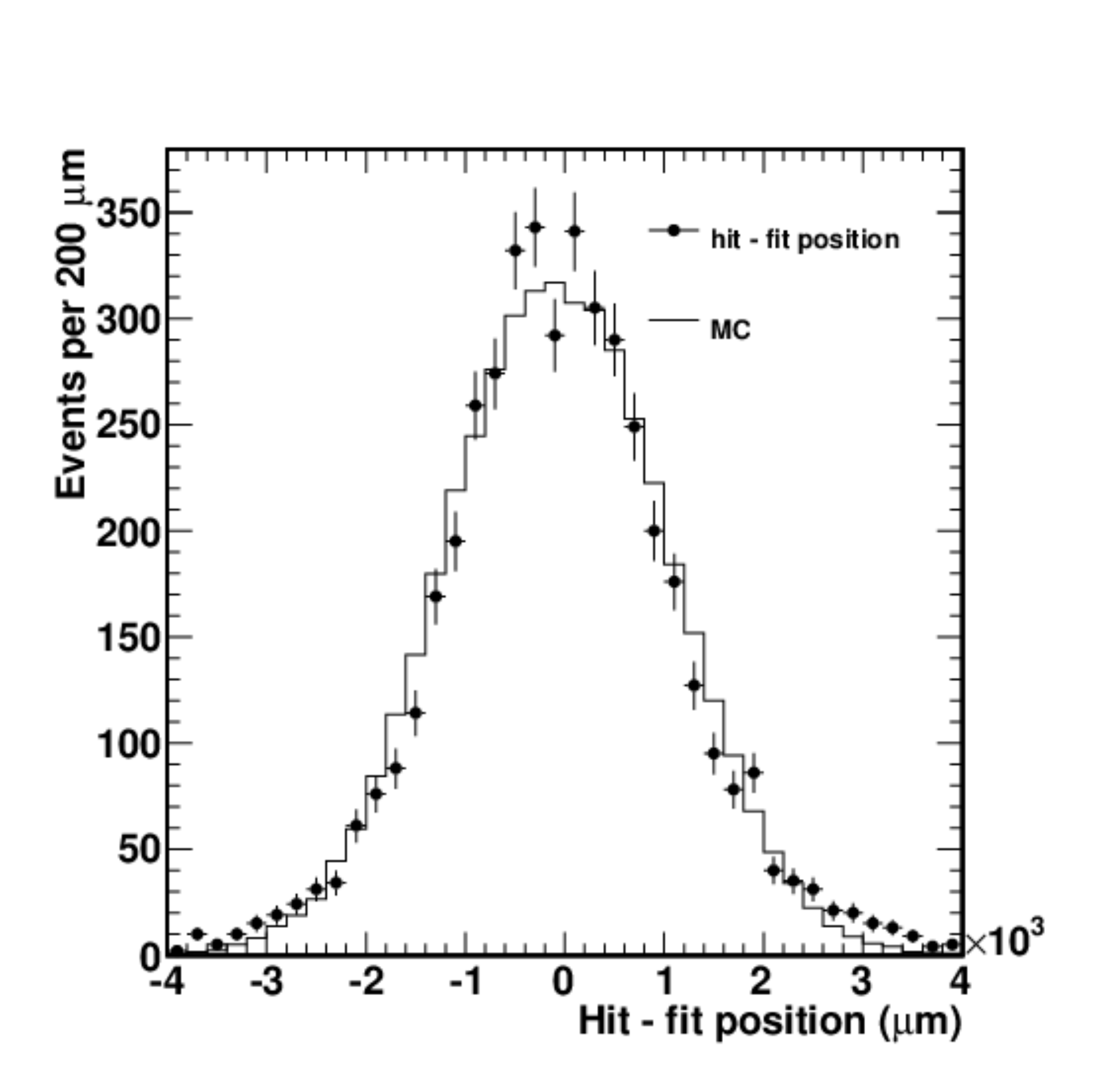}
\end{center}
\end{minipage}
\begin{minipage}{0.49\hsize}
\begin{center}
\includegraphics[width=0.9\textwidth]{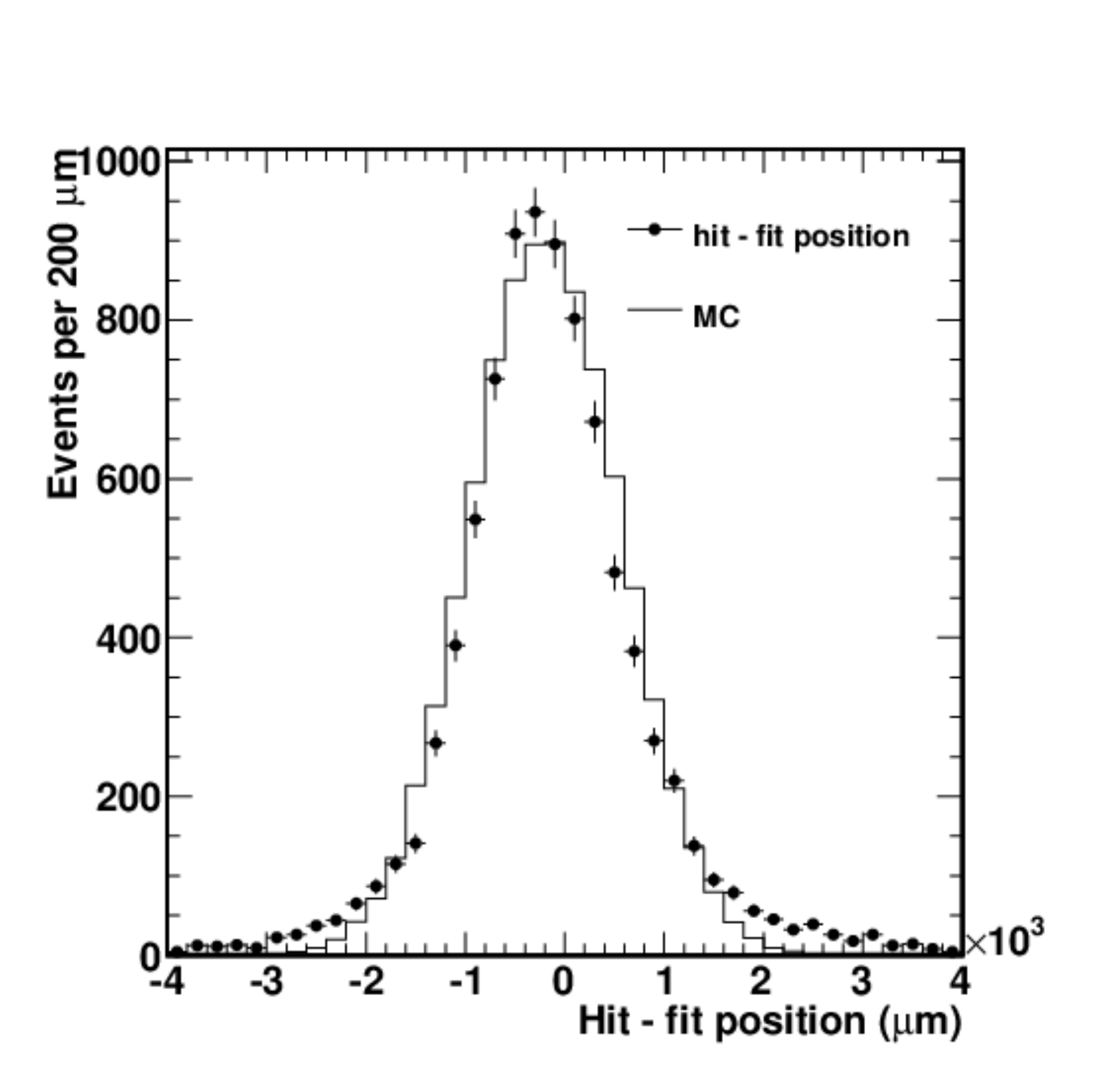}
\end{center}
\end{minipage}
\caption{\label{fig:posres}Histograms of position resolution for electron energies of $16$ MeV (left) and $30\ \mathrm{MeV}$ (right).}
\end{figure}

To estimate the position resolution, three MWPCs were installed with a spacing of $5\ \mathrm{cm}$ in series along the beam line. 
The difference between the hit position on the middle chamber and the expected position (fit position) estimated by the straight line connecting two hits found in the first and third chambers is calculated. Fig.~\ref{fig:posres} shows histograms of hit minus fit position for energies of $16\ \mathrm{MeV}$ and $30\ \mathrm{MeV}$. 
By fitting the histograms with a Gaussian plus constant, the standard deviations are found to be $(1050\pm 18)\ \mu\mathrm{m}$ and $(742.4 \pm 8.7)\ \mu\mathrm{m}$ for $16\ \mathrm{MeV}$ and $30\ \mathrm{MeV}$, respectively. \\
\indent
A simple simulation study is performed to reproduce the distributions for both beam energies in Fig.~\ref{fig:posres} considering effects of multiple scattering due to the materials of MWPC and air.
The results are included as histograms in the figure.
From the parameter of this simulation, the hit position resolution combining intrinsic position resolution of the MWPC and the analysis method is estimated to be $(640\pm 37)\ \mu\mathrm{m}$. 
Note that this measurement is strongly affected by the multiple scattering of beam electrons with rather low energies; thus, the observed resolution only shows the upper-limit of the intrinsic resolution. \\
%A simple simulation to find the difference between the fit position and hit position on the second MWPC was performed, considering the amplitude of multiple scattering due to the thickness of the MWPC and air at each energy. The standard deviation of resolution including the intrinsic position resolution of the MWPC and that of the analysis method was found to be $(640\pm5)\ \mu\mathrm{m}$. Please note that this measurement is strongly affected by the multiple scattering of low-energy electrons, and thus the observed resolution only show the upper-limit of the intrinsic resolution.\\
\indent
In the DeeMe experiment, four MWPCs and an electromagnet will be installed in the J-PARC MLF H1 area. The signal electron of muon-to-electron conversion has a monochromatic momentum of $\simeq 105\ \mathrm{MeV/}c$, and it is bent by a nominal angle of $\simeq 70^{\circ}$ with a curvature radius of $90\ \mathrm{cm}$ in the magnetic spectrometer. From these parameters, 
the momentum resolution solely due to the hit position resolution of $640$ $\mu$m would be approximately $0.1$ MeV/$c$,
which is obtained by simply propagating the hit position uncertainty to curvature estimation for a reasonable chamber configuration.
%the equipment will be put so that it can be bent $70^{\circ}$ with a radius of curvature of $90\ \mathrm{cm}$ in the magnetic spectrometer. Since the calculated momentum is proportional to the radius of curvature, using tracking detectors with a standard deviations of position resolution of $640\ \mu\mathrm{m}$ for a radius of curvature of $90\ \mathrm{cm}$ gives a momentum resolution of $0.07\ \mathrm{MeV/}c$. This satisfies the momentum resolution required for the experiment.
%
%==================================================
\section{Conclusions}
%==================================================
HV switching MWPCs for dynamical gas gain control have been developed. They were tested using the electron linac at Kyoto University Institute for Integrated Radiation and Nuclear Science. It was confirmed that the detector becomes 98\% efficient for detecting a single electron.
Position resolution was also estimated, and the contribution to the total momentum resolution was expected to be small
for the DeeMe experiment.
%
%The pulse height was obtained by hit finding by a method of summing up adc points in the cathode strip (channel) direction and the sample point (time) direction. These results confirm that the detector has a sufficient detection efficiency in the case of single electron. Position resolution measurements have also been done and that finds to meet the requirements of the momentum resolution for the setup of the equipment installation in the J-PARC MLF H1 area.
%
%
%==================================================
\section{Acknowledgements}
%==================================================
This work was supported by JSPS KAKENHI grant number JP24224006 and 17H01128. Part of this work was performed using facilities at Kyoto University Institute for Integrated Radiation and Nuclear Science. We thank the staff of the beam facilities for their support during test experiments, especially N. Abe and T. Takahashi of KURNS-LINAC.
%
%




\begin{thebibliography}{00}


%% \bibitem{}

%==================================================
\bibitem{SM}
%==================================================
S.~L.~Glashow, Nucl. Phys. {\bf 22}, 579 (1961);
S.~Weinberg, Phys. Rev. Lett. {\bf 19}, 1264 (1967);
A.~Salam, in {\it Elementary Particle Theory}, edited by N.~Svartholm (Almqvist and Wiksell, Sweden, 1968), p.367.
\bibitem{BSM} M. Raidal and A. Santamaria, Phys. Lett. B {\bf 421}, 250-258 (1998);
K. S. Babu and Cristopher Kolda, Phys. Rev. Lett. {\bf 89}, 241802 (2002).
\bibitem{MWPC} H. Natori {\it et al.}, Prog. Theor. Exp. Phys. {\bf 2017}, 023C01 (2017).
\bibitem{fadc} N. M. Truong, {\it et al.}, IEEE Transactions on Nuclear Science {\bf 65}, 9 (2018).
%\bibitem{Wire-Resonance} 
\bibitem{4} M. Morii, T. Taniguchi, and M. Ikeno, {\it Development of a readout electronic system for the VENUS vertex chamber}, KEK Internal 87-14 (1988).
\bibitem{5} N. Teshima, in {\it proceedings of the Flavor Physics \& CP Violation 2015 (FPCP2015) conference}, PoS (FPCP2015) 061 (2016).
\bibitem{HLine} N. Kawamura {\it et al.}, Prog. Theor. Exp. Phys. {\bf 2018}, 113G01 (2018).
\bibitem{Permittivity} Gases considered for the MWPCs have almost the same permittivities as the vacuum at a 0.1\% level.
\bibitem{Takezaki} Y. Takezaki, {\it HV discharge tests of MWPC wires and development of a HV control system for the MWPC used in the DeeMe experiment searching for muon-electron conversions}, Master's Thesis, Osaka City University, unpublished (2016).
\bibitem{Paschen} F. Paschen, {\it Annalen der Physik} {\bf 273} (5), 69-75 (1889).
\bibitem{3} CERN (12 February, 2018). {\it Garfield++ - simulation of tracking detectors}. Retrieved April 2, 2018, from http://garfieldpp.web.cern.ch/garfieldpp/
\bibitem{Linac}{The Linac actually produces 30 MeV electron beams in normal operation with higher beam intensities.}

\end{thebibliography}


\end{document}